\documentclass{aa}

\usepackage{graphicx}

\usepackage{txfonts}
\usepackage{xcolor}
\usepackage{threeparttable}
\usepackage[export]{adjustbox}

\usepackage{footmisc}
\usepackage[colorlinks=true,linkcolor=blue,debug=true]{hyperref}
\usepackage{comment}

\begin{document}

   \title{Detection of Oscillation-like Patterns in Eclipsing Binary Light Curves using Neural Network-based Object Detection Algorithms}


   \author{B. Ula\c{s}\inst{1,2},
          {T. Szklen\'{a}r}\inst{1},
          {R. Szab\'{o}}\inst{1,3}
          }

   \institute{Konkoly Observatory, Research Centre for Astronomy and Earth Sciences, HUN-REN, MTA Centre of Excellence, Konkoly-Thege Mikl\'{o}s \'{u}t 15\textendash{}17., 1121, Hungary\\
         \and
              Department of Space Sciences and Technologies, Faculty of Sciences, \c{C}anakkale Onsekiz Mart University, Terzio\v{g}lu Campus, TR~17100, \c{C}anakkale, Turkey\\
             \email{burak.ulas@comu.edu.tr}\\
         \and
         E\"otv\"os Lor\'and University, Institute of Physics and Astronomy, H-1117, Budapest, P\'azm\'any P\'eter s\'et\'any 1/a, Hungary
             }

   \date{Received September 15, 1996; accepted March 16, 1997}

 
  \abstract
   {}
   {The primary aim of this research is to evaluate several convolutional neural network-based object detection algorithms for identifying oscillation-like patterns in light curves of eclipsing binaries. This involves creating a robust detection framework that can effectively process both synthetic light curves and real observational data.}
  {The study employs several state-of-the-art object detection algorithms, including Single Shot MultiBox Detector, Faster Region-based Convolutional Neural Network, You Only Look Once, and EfficientDet besides a custom non-pretrained model implemented from scratch. Synthetic light curve images and images derived from observational TESS light curves of known eclipsing binaries with a pulsating component were constructed with corresponding annotation files using custom scripts. The models were trained and validated on established datasets, followed by testing on unseen {\it{Kepler}} data to assess their generalization performance.  The statistical metrics are also calculated to review the quality of each model.}
   {The results indicate that the pre-trained models exhibit high accuracy and reliability in detecting the targeted patterns. Faster R-CNN and You Only Look Once, in particular, showed superior performance in terms of object detection evaluation metrics on the validation dataset such as mAP value exceeding 99\%. Single Shot MultiBox Detector, on the other hand, is the fastest although it shows slightly lower performance with a mAP of 97\%. These findings highlight the potential of these models to contribute significantly to the automated determination of pulsating components in eclipsing binary systems, facilitating more efficient and comprehensive astrophysical investigations.}
   {}


\titlerunning{Detection of Oscillations in EBs using CNN-based algorithms}

   \maketitle

%

\section{Introduction}

Eclipsing binaries with pulsating components (EBPCs) are one of the key stellar system classes that reveal the physics behind pulsating stars since their mass can be directly measured. They show pulsation patterns in their characteristic eclipsing binary light curve shapes, thus the oscillation and binary properties can be uncovered by separate analyses of these effects. The variety of the pulsation class and spectral types allows researchers to understand the oscillation phenomena in different circumstances and to improve the understanding of binary systems. Recent studies show that the number of these kinds of systems is increasing, mainly thanks to accurate space-based missions like Transiting Exoplanet Survey Satellite \citep[TESS,][]{ric15} and {\it{Kepler}} \citep{bor10}. Some specific classifications, especially systems with $\delta$~Sct components, were made by \citet{pop80} and \citet{and91} for detached binaries as EA/DSCT and oEA (oscillating Eclipsing Algols) were proposed for semidetached systems by \citet{mkr02,mkr04}.

Machine learning applications have taken place in our lives expeditiously, as have scientific investigations in recent years. Computer vision tasks, as a significant subclass of machine learning, reflect its role in automating and enhancing visual data interpretation in many disciplines. Inevitably, the use of machine learning methods increased steeply in astrophysics as it opened a new window to evaluate the data taken from the sky. Classifying and characterizing celestial objects is one of the most common uses of machine learning in astronomy, besides simulating astrophysical phenomena with better performance, as \citet{sza22} remarked. The processes can be automated by researchers using labeled datasets containing information about various types of stars, galaxies, supernovae, and other objects. This implies a serious reduction in the time and effort needed for manual analysis. In addition, astronomers can now discover objects that would have been missed using conventional approaches and machine learning techniques can deepen our grasp of the diversity and complexity of the universe.

Specific to binary stars, researchers used machine learning and deep learning techniques to detect, fit, and classify the light curves of binary systems. \citet{wyr03} used the OGLE data \citep{uda98} to identify 2580 binary systems in the Large Magellanic Cloud by proposing an artificial neural network approach. \citet{prs08} introduced an artificial neural network trained using 33235 detached eclipsing binary light curve data to determine some physical parameters. \citet{koc20} discussed several fitting techniques and concluded that the initial parameters of the binaries can be estimated with the help of machine learning techniques. An image-based classification of variable stars using machine learning methods was done by \citet{szk20} where a class for eclipsing binaries was also included. A two-class morphological classification of \citet{cok21} focused on different deep-learning methods, including convolutional neural networks (CNNs) based on synthetic light curve data. \citet{bod21} set a machine learning algorithm using a locally linear embedding method to classify the morphologies of OGLE binaries. \citet{szk22} classified the variable stars, comprising the eclipsing binaries, based on their visual characteristics using a multi-input neural network training with OGLE-III data. In a recent study, a classification of more than 60000 eclipsing binaries was made by \citet{hea24} using machine learning algorithms based on The Zwicky Transient Facility data.      

\section{Overview and Methodological Approach}\label{sec:overview}

The purpose of this study is to employ CNN-based object detection models to train the machine and to detect patterns similar to light variations arising from the pulsations of one/both of the components in eclipsing binaries using their light curves. We first constructed the appropriate light curve images and then developed models trained using several detection algorithms, with the goal of identifying pulsations and minimum patterns in any given image. Since the methods applied in this study are more efficient than manual detection, we intended to offer promising new ways for identifying such patterns and discovering new members of the EBPC zoo.

The training and detection processes in the present study were designed to be applied to light curve images prepared initially to facilitate the application of the algorithms and achieve the study's objectives. EBPCs exhibit two characteristic patterns in their light curves: light variations due to pulsation and occultation. Therefore, each light curve image in the training and validation datasets must include at least one instance of both patterns. This fact highlights the importance of carefully selecting the time interval used in image construction. Since pulsation variations are most prominent during maximum phases, the interval should include at least one of the light curve maximums in addition to the minimum. Furthermore, the time interval plays a crucial role in ensuring the clear visibility of pulsation patterns in the images. If the interval is too long, the fine details of the oscillations may be lost and become indistinguishable. Conversely, if the interval is too short, there is a risk that one of the characteristic patterns may be missing from the image. To address this in our investigation, we constructed light curves using a time interval of 0.7 times the orbital period of a given binary system, a value determined after several iterations. This approach ensures {\it{(i)}} that both a pulsation and an eclipse pattern are present in the image, and {\it{(ii)}} that the image adequately covers the pulsation region.

In the present study, pattern detection for the validation dataset was conducted based on the Intersection over Union (IoU) and confidence score parameters whereas the latter is the indication of the detection quality for the test set formed by unseen data. The confidence score, ranging from 0 to 1, reflects the model's estimated probability that a detected object belongs to a specific class, thereby corresponding to the likelihood that the predicted bounding box accurately contains the object of interest. \citet{red16} remarked that the confidence indicates the model's certainty that a bounding box contains an object and its accuracy of the box's predicted location. The selection of a threshold often depends on the task \citep{lin17} and is typically refined through iterative trials. In our investigation, we set the confidence threshold to 0.5, a value chosen after several trials, which helped us minimize the impact of false positives and false negatives (Sec.~\ref{sec:keptest}). IoU, on the other hand, measures the accuracy of a predicted bounding box relative to the ground truth. It is calculated by dividing the area of overlap between the predicted bounding box and the ground truth box by the area of their union.  We calculated resulting metrics for various IoU thresholds for a versatile investigation of the models' performances, while a threshold of 0.5 was adopted for generating the confusion matrices, meaning that the overlap must be at least 50\% for a detection to be considered correct following \citet{eve10}. To restate our primary objective, we aim to accurately identify whether a given system is a potential EBPC by detecting the presence of specific patterns in its light curve. It is worth emphasizing that the exact location of the pulsation patterns within the image is not critical; as long as they are detected with the defined level of confidence in any region, the system can be considered a potential EBPC.

In the following section, we explained the properties and preparation process of the data used in the study. Sec.~\ref{sec:detection} deals with our efforts in training the models and detecting patterns of interest on the light curve images using five different CNN-based algorithms. A test of detecting pulsation and minimum patterns on binary systems from another database, short cadence data from {\it{Kepler}} mission, is presented in Sec.~\ref{sec:keptest}. We summarized the results, gave the concluding remarks and drew the future perspective in the last section.

\section{The Data}\label{sec:data}

\subsection{Data augmentation}\label{subsec:augm}
The models used in this study are fed with data (light curve images and corresponding annotation files) to be trained and to detect regions of interest in the images. At first, the idea of constructing images from known EBPCs may seem brilliant. However, the number of known systems of this type keeps the dataset small and consequently avoids reaching the desired precision in detecting patterns using neural network algorithms. Therefore, to increase the amount of training and validation data, we constructed synthetic light curve images and corresponding annotation files that mimic the observational data. Combining them with the small amount of observational light curves allows us to achieve a considerable amount of data to train a machine using certain algorithms.

\subsection{Synthetic data}\label{subsec:syn}
Synthetic light curve images were constructed by applying the following procedure: {\it{(i)}} Determination of parameter intervals for detached and semidetached binary systems and construction of eclipsing binary light curve data using the derived intervals. {\it{(ii)}}  Applying random noise, adding artificial oscillations on the light curve data and then constructing light curve images and annotation files. {\it{(iii)}}  Shifting images in horizontal and vertical directions randomly by keeping the bounding box area to make the data random and diverse.

\subsubsection{Determination of Parameter Intervals}\label{subsec:par_det}

Since almost all pulsating components in eclipsing binaries are observed in detached and semidetached systems, we decided to start by constructing synthetic light curves of those types of binaries. The parameters used in the light curve construction play an important role in obtaining realistic results and, more importantly, maintaining physically meaningful boundaries. Therefore, we first calculated the key light curve parameters (mass ratio, inclination, and surface potentials of the components) based on the conventional formulae and approximations (see Appendix \ref{sec:appa}) for eclipsing binaries using the absolute parameters of 212 detached and 110 semidetached systems provided by \citet{sou15} and \citet{mal20}, respectively. Alongside the orbital period and effective temperature values presented in these studies, we determined the value intervals for the seven essential parameters (Table~\ref{tab:interval}) needed to construct accurate synthetic light curves. This approach enabled us to define realistic bounds for detached and semidetached binaries.

\begin{table}
    \caption{Parameter intervals for detached (D) and semidetached (SD) systems used in synthetic data construction. $P$, $i$ and $q$ refer to the orbital period, inclination and mass ratio of the system, respectively, while ${T_{e_1}}$, $T_{e_2}$, $\Omega_1$ and $\Omega_2$ denote the effective temperatures and surface potentials of the components.  The potentials of the secondary components are assumed to be equal to their critical value, $\Omega_{cr}$, in semidetached systems.}
    \label{tab:interval}
    \centering
    \begin{tabular}{lcc}
        \hline
         Parameter   & \multicolumn{2}{c}{Value Range} \\
         & D& SD \\
        \hline    
        $P~(\mathrm{days})$  & [0.452, 12.426]& [0.525, 7.160]\\
        $i~(\mathrm{degrees})$  & [63, 85]& [58, 75]\\
        $q$  & [0.493, 1.186]& [0.092, 0.712]\\
        $T_{e_1}~(\mathrm{K})$  & [2992, 13002]& [5000, 25200]\\
        $T_{e_2}~(\mathrm{K})$  & [2999, 10990]& [3900, 12690]\\
        $\Omega_1$ & [2.729, 20.088]  &[2.138, 7.870]\\
        $\Omega_2$ & [3.461, 20.885]   &$\Omega_{cr}$\\
        \hline
    \end{tabular}
\end{table}

\begin{figure*}
   \centering
   \includegraphics[width=\textwidth]{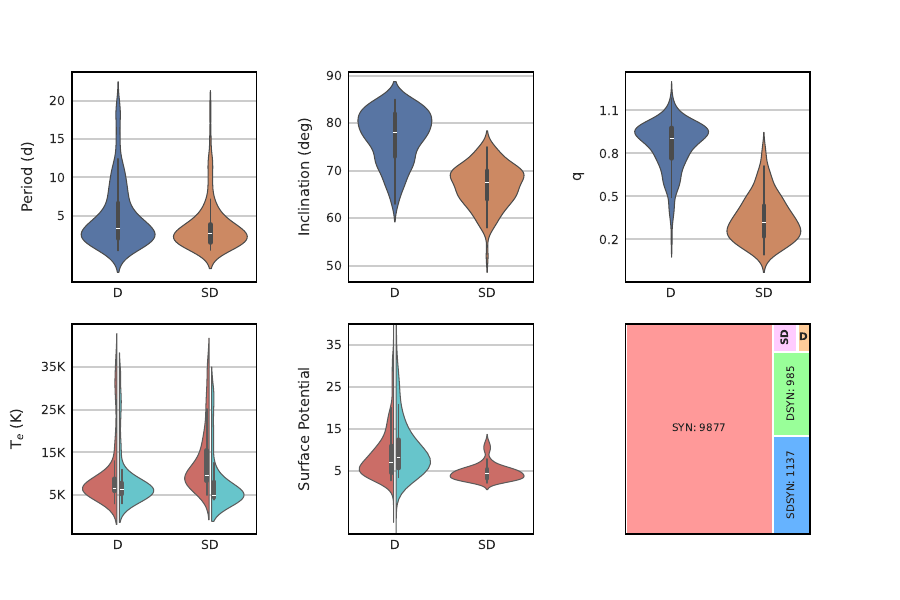}
   \caption{Violin plot showing the distribution of the parameters used in synthetic construction of light curves for detached (D) and semidetached (SD) binaries. The white lines at the centers are the median values covered by black boxes whose lower and upper boundaries are first and third quartiles, $Q1$ and $Q3$ \citep{krz14}. The two ends of the lines overlap the boxes, denoting the minimum and maximum values of the parameter. The density plots around the boxes represent the probabilities of the corresponding values in the y-axis, where the wider refers to more frequent occurrences. The effective temperature ($T_{e}$) and surface potential distributions for the primary (red, left) and secondary (green, right) components are shown in split areas for the sake of a smooth comparison of the values in the two leftmost plots at the bottom pane. We note that the surface potential plot for the semidetached systems is only values for components well detached from their Roche lobe, since the potential of the other component is assumed to be equal to its critical value, indicating that the component fills its Roche lobe. The last box is a treemap plotted based on the number of light curves in different phases of light curve construction. The numbers of detached and semidetached systems from catalogues are 110 and 212 (two small tiles in the upper right). DSYN and SDSYN are synthetic detached and semidetached light curves constructed using the catalogue data while SYN refers to the total number of synthetic light curves mimicking EBPCs and used in model training and validation.}
              \label{fig:boxplot}%
    \end{figure*}

Once the parameters and the intervals were derived, we managed to construct 985 detached and 1137 semidetached synthetic light curves by employing two of our scripts\footnote{\url{https://github.com/burakulas/code_repo/}\label{fn:repo}} which automate the Wilson-Devinney method \citep{wil71,wil20} and sets various inclination values together with each parameter group as input parameters. To clarify the procedure, for instance, the maximum value for the inclination of detached binaries is 85$\,\mathrm{degrees}$, as seen from Fig.~\ref{fig:boxplot}. We derived several inclination values for a given parameter set (inclination, orbital period, mass ratio, effective temperatures, and surface potentials) by increasing the initial inclination by 1.8$\,\mathrm{degrees}$ until 85$\,\mathrm{degrees}$. To illustrate, based on the above logic, an initial inclination value of 79$\,\mathrm{degrees}$ would yield four inclinations: 79, 80.8, 82.6, and 84.4$\,\mathrm{degrees}$. Then we constructed light curves implementing each inclination value with the other six parameters. Therefore, we obtained 1078 synthetic light curve data (phase-flux pairs) in detached configuration. The number of synthetic semidetached light curves was 1205 using the same method and a lower increment in inclination.  However, the resulting number of constructed light curves decreased since the process needed a post-check to eliminate the potentially unphysical light curves. The elimination was done in two steps: first by scanning the resulting data for unusual values (negatives, zeros, error messages, etc.) and then by inspecting the shape of the constructed light curves via eye by plotting them to a file to be sure that the data is proper. The systems with orbital periods larger than 20$\,\mathrm{days}$ were not included in our calculations for both configurations, besides the parameters of outliers among semidetached binaries having a mass ratio larger than 1.0 were excluded. The eccentricities are assumed to be zero as they are not distinctive from our aim of detecting patterns of interest. The distribution for each parameter used in the light curve construction procedure is given in Fig.~\ref{fig:boxplot} along with a treemap showing the number of light curves constructed in various parts of the data preparation. 

\subsubsection{Superposing the Artificial Oscillations}\label{sec:suppos}

\begin{figure*}
   \centering
   \includegraphics[width=\textwidth]{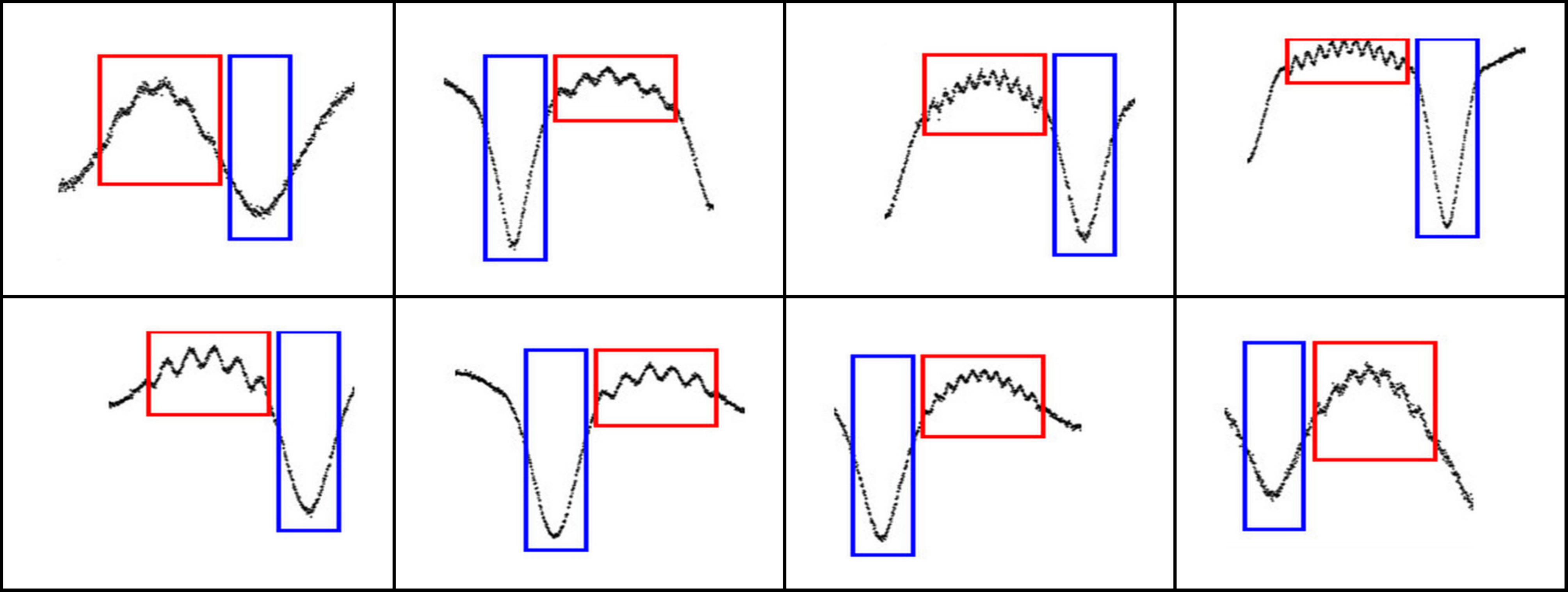}
   \caption{Samples from the synthetic light curve images with the ground truth bounding boxes corresponding to pulsations (red) and minimum (blue) patterns.}
              \label{fig:syndata}%
\end{figure*}

The light curves yielded using the method explained in the previous subsection show characteristic binary star light curves without including any pulsation pattern. To resemble the observational light curves of EBPCs, oscillation effects must be added. In this process, it is important to assume the location of the patterns, as the algorithm accepts input images along with the corresponding annotation files that indicate the location of the pattern in question. We applied light variations mimicking the oscillations to our data using the script\footref{fn:repo} written for this aim. The code adds random noise to the light curve data remarked in the previous subsection and then superposes a pulsation pattern on the maximum I (phases between 0.1 and 0.4) based on six values entered by the user. The first three are the number of cycles that are to be seen in a 0.3-phase interval. The rest are initial and final amplitude values and an amplitude increment in magnitude unit. The code converts the cycle numbers to frequencies in d$^{-1}$ based on the orbital period value, constructs oscillation-like light variation using the frequencies and amplitudes, locates the light variation on the binary light curve simulating the light curves of EBPCs, and produces images. It also exports an annotation file, obligatory to conduct the detection algorithm, for each image referring to the coordinates of ground truth bounding boxes (x-center, y-center, width, height) which can be easily converted to Pascal VOC XML and YOLO format. The bounding boxes correspond to two patterns of interest, pulsation and minimum, since we deal with the pulsations in eclipsing binaries and aim to see both patterns to be sure that the light curve belongs to an EBPCs. Although the procedure so far made us very close to starting training with the constructed data, we applied two more steps: random shift and vertical flip. Random shift avoids the patterns of interest squeezing in similar regions on each image. The shift causes the target patterns to be located in images randomly and thus leads the synthetic curves to resemble the observations more, besides overcoming the location bias that may arise from the artificial oscillations occurring at the same phases. This was done by a Python code\footref{fn:repo} using the appropriate shift parameters that do not allow the patterns to be out of the image boundary. Applying a random vertical flip (left-right flip) to randomly selected images increases the chance of mimicking the observational light curves, where pulsations are observed before the primary minimum. Making the data as random as possible is significant to avoid skewing the conclusions. After a last check and elimination by the eye, as done to the light curves in the previous subsection, a total of 9877 synthetic light curve images\footnote{\url{https://drive.google.com/file/d/1nmcpvd1IWXL_BqgFGajEQO4fNxSq6G8I/view?usp=sharing}\label{fn:lcdata}} were decided to be used in detection models. Eight samples from resulting images with their annotations are shown in Fig.~\ref{fig:syndata}. One might consider that the whole process seems too tedious and obsessive to be proper to the human eye which extends the time spent for yielding the data, however, it is known that the data preparation, especially human-confirmed ones, is the most time-consuming part of machine learning studies as pointed out by \citet{abd17}. The data is also worth going under the mentioned re-checks regarding its effect on the performance of a computer vision project.

\subsection{Observational data}\label{sec:obsdata}

We collected known EBPCs from four main studies; \citet{lia17}, \citet{ali22}, \citet{shi22}, \citet{ali23} in which the first two and last one lists the systems with $\delta$~Sct type components while the third one catalogues the pulsators in EA-type binaries. The total number of the systems reached 426 after a cross-match based on the target name. The light curve image for a given binary was constructed through a series of steps utilizing multiple code snippets. The process began with reading target names from a predefined list of known EBPCs and accessing the Barbara A. Mikulski Archive for Space Telescopes (MAST) database to verify the existence of data files corresponding to the targets. For targets that passed the first step, we performed a target name confirmation process by inspecting the headers of the FITS files to ensure the data belonged to the intended targets. Next, we extracted the TIME and SAP\_FLUX values from the FITS files, removing any invalid or NULL data points. The flux values were then converted into magnitudes ($m_i$) using the formula $m{_i} = -2.5logF{_i}$, where $F{_i}$ represents the SAP\_FLUX values in the corresponding FITS files. To determine the time intervals for each light curve, the orbital period values from the TEBS catalog \citep{prs22} were multiplied by a user-defined period factor (in our case, 0.7, see Sec.~\ref{sec:overview}). This ensured that the time-magnitude pairs for each system were tailored to our analysis objectives. Finally, the light curve images were generated by plotting the time-magnitude pairs. The patterns for those images are annotated by hand using \textsc{labelImg}\footnote{\url{https://github.com/HumanSignal/labelImg}} software. The light curves showing extreme scattering and the ones with unclear patterns were eliminated. We achieved 330 images from observational data through this process, however, the number is still quite small for our aim. 

\begin{figure*}
   \centering
   \includegraphics[width=\textwidth]{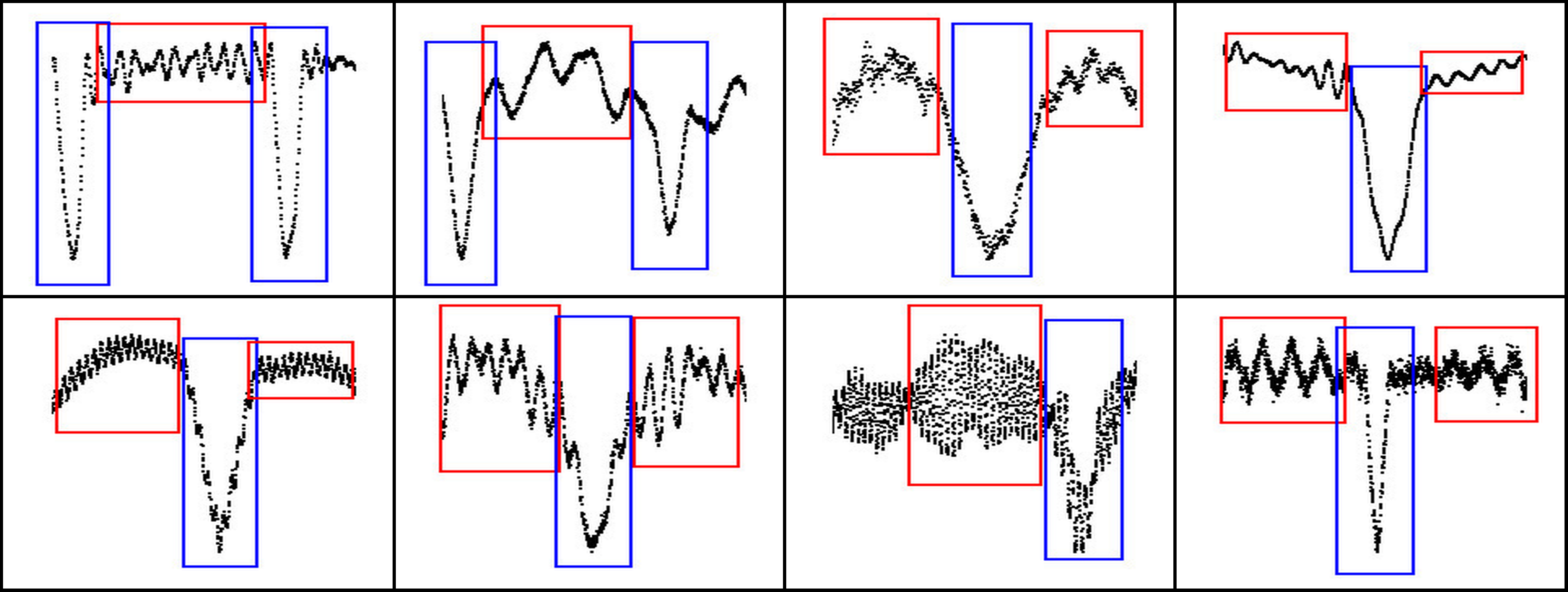}
   \caption{Same as Fig.~\ref{fig:syndata}, but for observational light curve images as exported from \textsc{DetOcS}.}
              \label{fig:obsdata}%
\end{figure*}

An approach that we conducted to increase the number of observational data was dividing the light curve of each target into several parts in time by using a certain factor of the orbital period. For instance, seven fits files from the TESS mission meet the conditions when applying a search for the system TIC~48084398 in the Barbara A. Mikulski Archive for Space Telescopes (MAST) portal. The observation durations ($T_{stop}-T_{start}$) of the data change from 24.86 to 28.20$\,\mathrm{days}$ corresponding to a total of 187.04$\,\mathrm{days}$ long. Dividing the complete observations into equal parts covering 0.7 of the orbital period in time ($0.7\times2.1218069=1.48\,\mathrm{days}$, in the case of TIC~48084398) results in obtaining almost 126 individual light curve data for a single system. This method led us to construct thousands of images yet annotating by hand is not effective anymore. Therefore, we developed a web implementation, \textsc{DetOcS}\footnote{\url{https://github.com/burakulas/detocs}\label{fn:detocs}} ({\bf{Det}}ection of {\bf{O}}s{\bf{c}}illations in Eclipsing Binary Light Curve{\bf{S}}), which reads the fits file of the corresponding target on MAST portal by using Astropy \citep{ast13, ast18, ast22}, divides the light curve data in time relying on the parameter given by the user, constructs light curve images, apply detection and create annotation files according to the detection results by using iterative object detection refinement method. The implementation also allows fast human confirmation of the detected patterns with one click, considerably easing the data collection procedure. \textsc{DetOcS} contains the above-mentioned code snippets as functions and also requires a pre-trained model. Therefore, we first create a TensorFlow object detection model trained with our synthetic (See~\ref{subsec:syn}) and 330 observational light curves by using the most computationally inexpensive model similar as explained to one that in Sec.~\ref{sec:ssd}. While we provide details in the next section, briefly summarizing here, when training the object detection model, the network learns to detect and localize objects in images by predicting bounding boxes and class labels in a single pass. In this way, we gathered 1493 observational light curve images with their annotations\footref{fn:lcdata}, more than 4.5 times the initial number. The eight samples directly taken from the output of {\textsc{DetOcS}} are shown in Fig.~\ref{fig:obsdata}.


\begin{figure}
   \centering
   \includegraphics[width=0.5\textwidth]{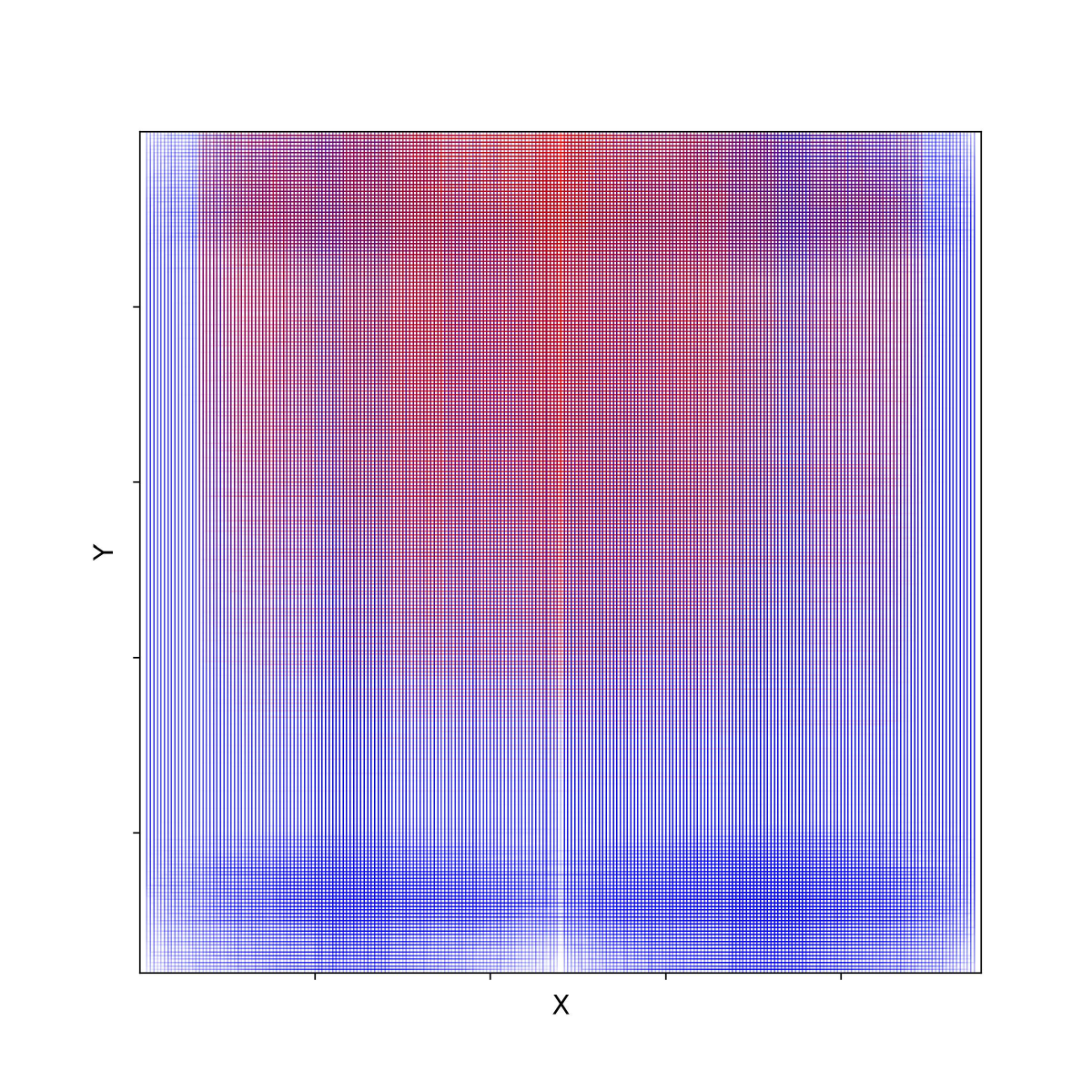}
   \caption{Plot of all pulsation (red) and minimum (blue) ground truth bounding boxes in one image area. See text for details.}
              \label{fig:annodist}%
\end{figure}

In conclusion, the total number of human-confirmed synthetic and observational light curve image data is 11370 and can be accessed online\footref{fn:lcdata} with their annotation files in Pascal VOC format. Fig.~\ref{fig:annodist} presents all the bounding boxes gathered into a single image, highlighting the effective distribution of annotations across the image area. The minimum and pulsation patterns achieve coverage of 98.3\% and 71.5\% of the area, respectively. The dense appearance of the red boxes on the upper part is not unexpected since the pulsation patterns are generally observed at maximum phases. For the reader who gets curious about why the confirmation by a human is crucial even for the observational data, we would like to remind you of the aim of the present computer vision study: to give the machine the ability to see the patterns in the light curves at least close to an experienced human does. Therefore, the careful preparation of all images and annotations in datasets is inevitable to succeed. 

\section{Detection of the pulsation and binarity}
\label{sec:detection}

In the present study, we employed CNN-based object detection models processing input images by first passing them through a series of convolutional layers, where they automatically learn hierarchical features at different spatial resolutions. A model extracts relevant patterns, which are essential for identifying objects. Once these features are obtained, the network generates candidate bounding boxes around potential objects in the image. Each bounding box is then classified into predefined object categories, and the model refines the predictions by adjusting the bounding box coordinates to more accurately localize the objects. The final output consists of class labels and their corresponding bounding box coordinates, providing a precise identification and localization of objects within the image.

We used four conventional and one custom object detection model during the training and detection phases based on two patterns, pulsation and minimum, the crucial patterns seen in the light curve of an eclipsing binary with the pulsating component. The annotation files for the images (see Sec.\ref{sec:suppos} and \ref{sec:obsdata}) which determine the location and size of the patterns of interest were converted to PASCAL VOC \citep{eve10} XML format excluding the YOLO model which needs a specific YOLO format consisting of center positions, widths and heights of the ground truth boxes different from XML format which bases on horizontal and vertical coordinates of the box corners. 

The 30\% of all images constructed from synthetic and observation light curves were split as validation dataset. Thus, our training set consists of 7959 images while the number of images in the validation dataset is 3411. One must keep in mind that using the following models on a custom dataset is a pretty complicated procedure including countless times of testing the related parameters and hyperparameters in algorithms. We summarized these procedures in the following sections by 
giving the results. Additional information about models and the crucial parameters of the training procedures can be found in the Appendix \ref{sec:appb}.

\subsection{Single Shot Multibox Detector Model}\label{sec:ssd}

Single Shot MultiBox Detector \citep[SSD,][]{liu16} combines the efficiency of multiscale feature maps and bounding boxes to detect various sizes of objects. The model uses feature maps in different sizes, which allow the detection of high-level and fine-grained information on images. It can detect multiple objects in various categories with relatively high accuracy, and therefore, is a favorite model for computer vision tasks.

\begin{figure*}
   \centering
   \includegraphics[width=0.8\textwidth]{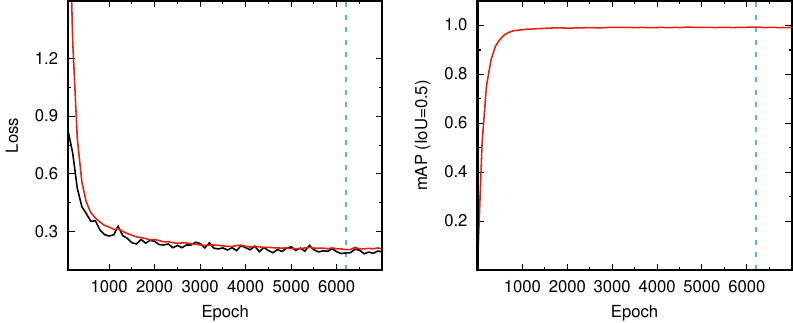}
   \caption{Loss and mAP variation for the SSD model during training and evaluation. The black and red lines refer to training and validation values, respectively. The blue dashed line corresponds to the 6200th epoch, the step for the final model. The plot data was smoothed exponentially by a factor of 0.6 to emphasize the trend. The TensorBoard parameter plots are available in the provided Google Colab notebook\protect\footnotemark{}.}
   \label{fig:SSD_loss_map}%
    \end{figure*}

\begin{figure*}
\centering
\begin{tabular}{ |c|c| }
\hline
\includegraphics[width=0.35\textwidth]{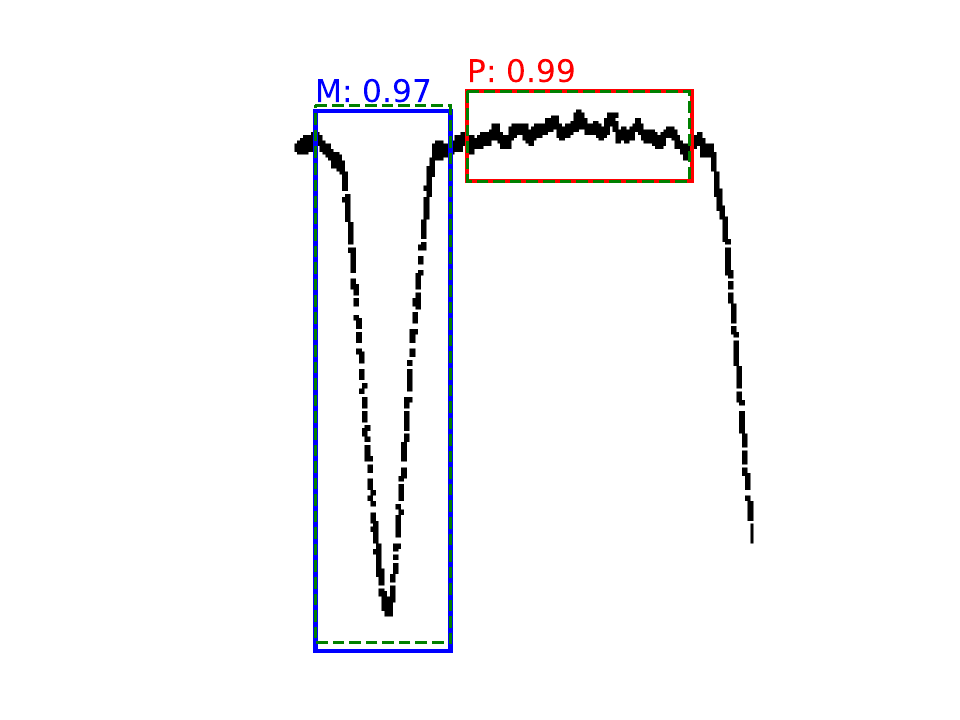} & \includegraphics[width=0.35\textwidth]{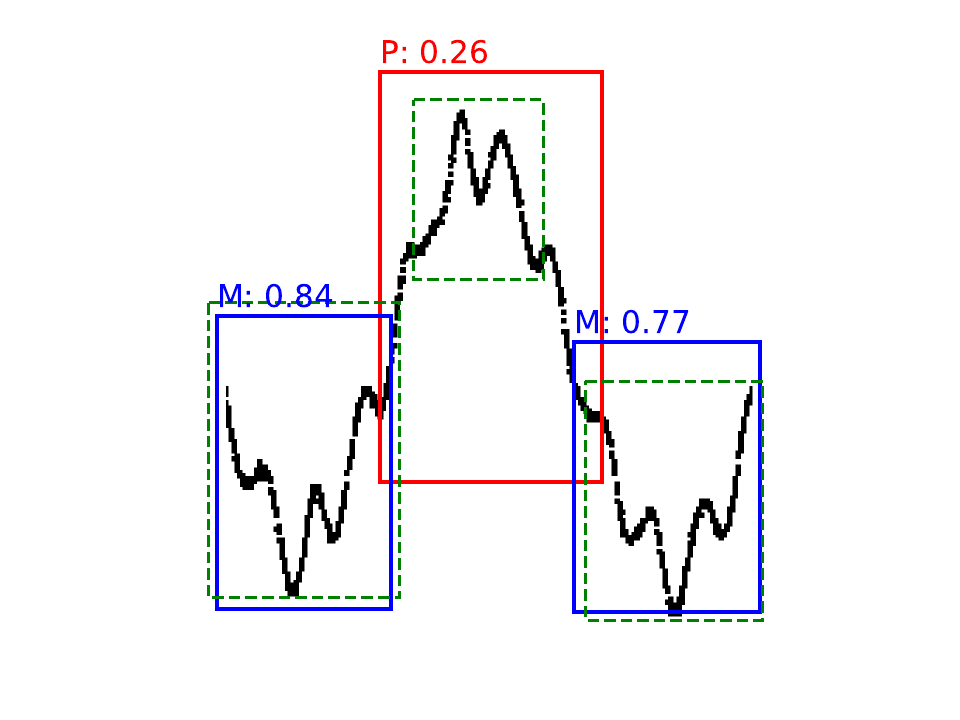} \\
\hline
\end{tabular}
\caption{Samples of different detection performances on observational light curve images in validation set having predicted bounding box for each ground truth annotation for the SSD model. Detected patterns (in red and blue bounding boxes) with the maximum (left, TIC~73672504) and minimum (right, TIC~232637376) average IoU values are compared to ground truth annotations (green dashed boxes). The numbers above the boxes correspond to IoU value for the related detection. P and M refer to pulsation and minimum patterns, respectively.}
 \label{fig:SSD_min_max}
\end{figure*}

\begin{figure*}
\centering
\begin{tabular}{ |c|c| }
\hline
\begin{minipage}{0.35\textwidth}
            \centering
            \includegraphics[width=0.9\linewidth]{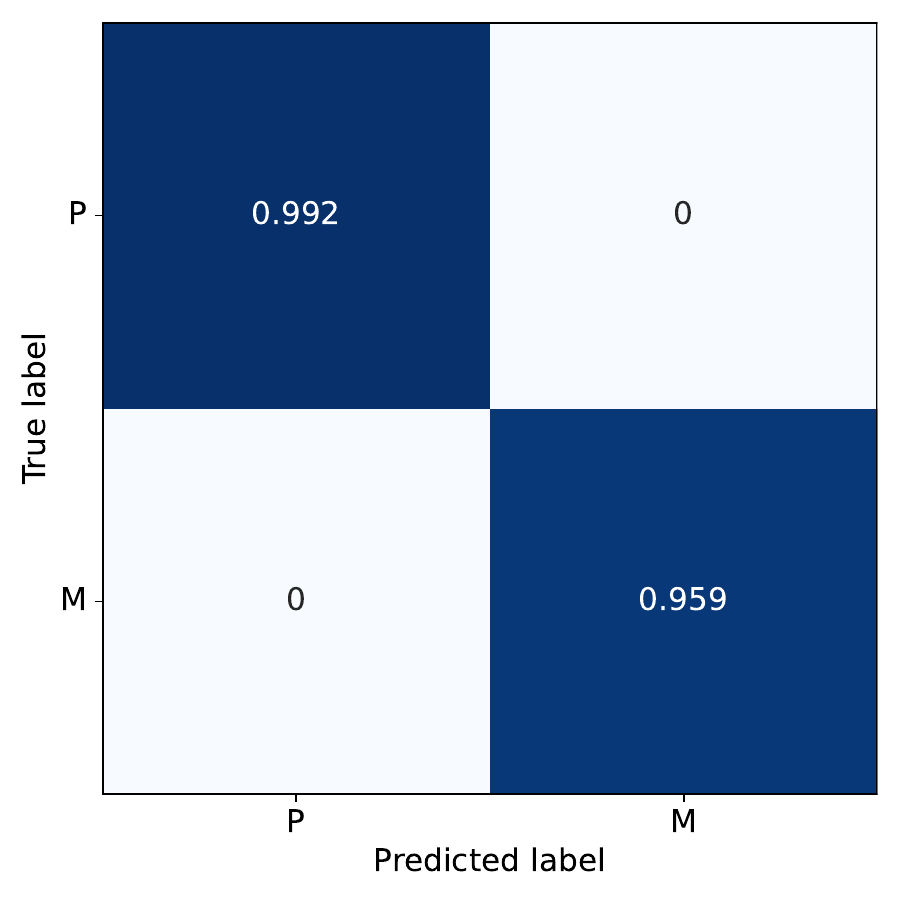}
        \end{minipage}
        &
        \begin{minipage}{0.35\textwidth}
            \centering
            \includegraphics[width=0.9\linewidth]{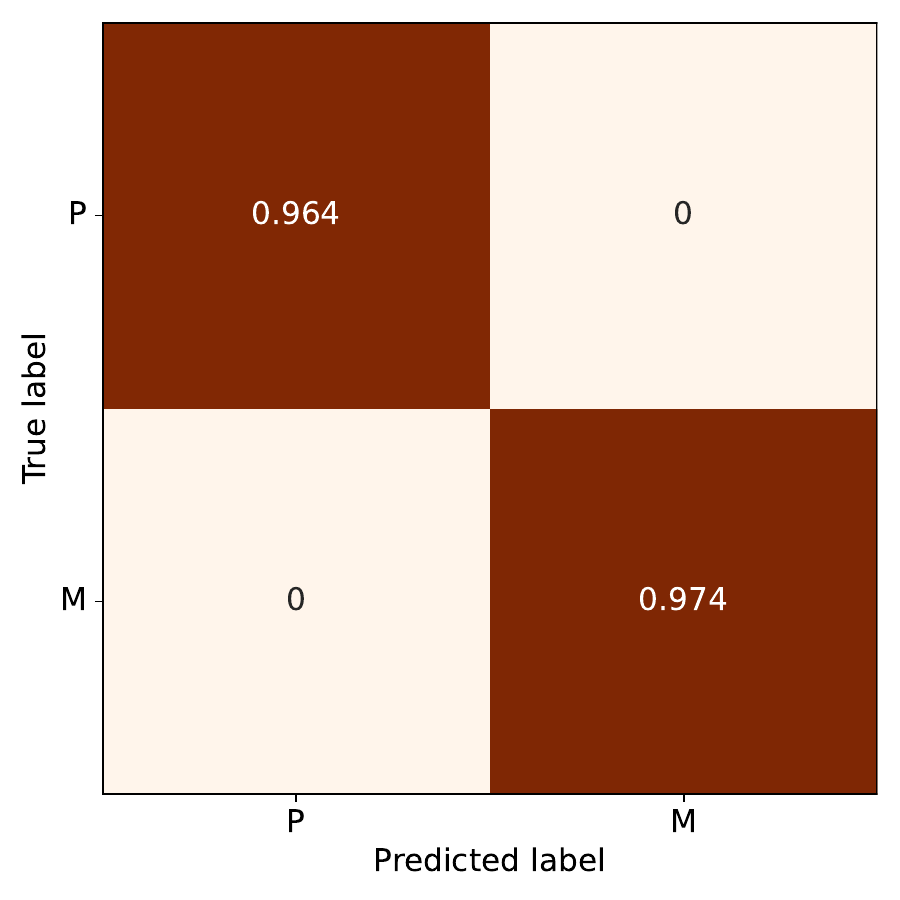}
        \end{minipage} \\

\hline
\end{tabular}
\caption{The confusion matrices for overall (left) and only observational (right) data in the validation dataset are calculated based on predicted class and bounding boxes by setting the IoU and confidence level to 0.5. We note that 0.08\%  of pulsation and 0.41\%  of minimum patterns are detected as background, namely, predicted as they are not objects of interest for overall data while the percentages are 0.36\% and 0.26\% for observations. P and M refer to pulsation and minimum patterns, respectively.}
 \label{fig:SSD_cfm}
\end{figure*}

\begin{figure*}
\centering
\includegraphics[width=\textwidth]{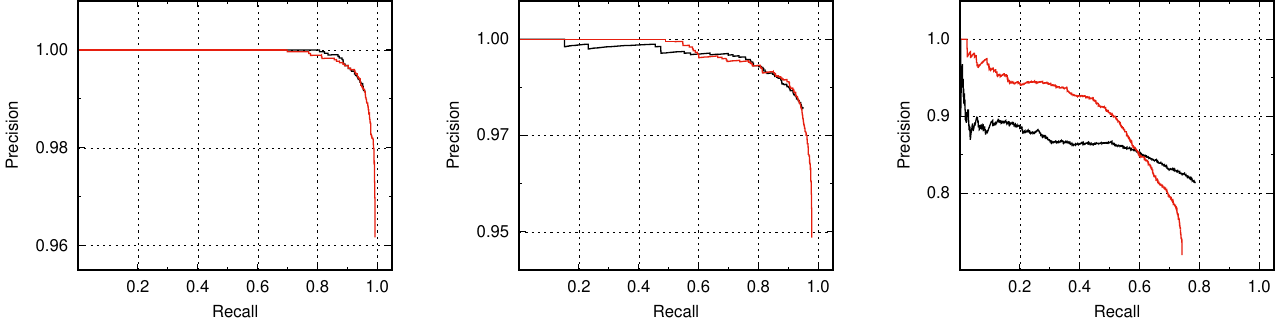} 
\caption{Precision - Recall curves of the detection in IoU=0.5 (left), IoU=0.75 (middle) and IoU=0.9 (right) thresholds. The black and red lines correspond to minimum and pulsation patterns, respectively. The calculations were conducted by using the code given by \citet{pad21}.}
 \label{fig:SSD_pr}
\end{figure*}

The object detection algorithm using the SSD method was set in the TensorFlow platform, specifically in TensorFlow Object Detection API\footnote{\url{https://github.com/tensorflow/models/tree/master/research/object_detection}\label{fn:tfapi}}. The model\footnote{\url{https://github.com/tensorflow/models/blob/master/research/object_detection/g3doc/tf2_detection_zoo.md}\label{fn:tfzoo}} forms in SSD algorithm using MobileNet~v2 \citep{san18} feature extractor. We first trained the model for 30$\,\mathrm{minutes}$ 7$\,\mathrm{seconds}$ until the 25000th epoch and observed the detection performance, which was unsatisfactory. Then the model from the last epoch was set as a new fine-tuning checkpoint and re-trained to achieve more quality detections. The re-training process took 14$\,\mathrm{minutes}$ and 13$\,\mathrm{seconds}$ until the 6200th epoch. Once the training was complete, we extracted a model file{\footnote{\url{https://drive.google.com/drive/folders/1aRu-PPFaLgPqAuuXIw5l8r6-0RYUZ7pC?usp=sharing}\label{fn:models}}}, thus we could test the performance of the model on the observational light curves in the validation dataset in detecting the patterns of interest. Using a final model file, the detection takes 0.22$\,\mathrm{seconds}$ per light curve image. The variation of training and validation loss functions through epochs are presented in Fig.~\ref{fig:SSD_loss_map} with the mean average precision at 0.5 IoU threshold, mAP (IoU=0.5). Both training and validation losses decrease steadily as training progresses, which is a positive sign indicating that the model is learning from the training data and improving its performance on the validation set. Continuous decrement in the validation set suggests that the model has learned the patterns in the data well and is generalizing effectively to the unseen data. A low training and validation loss also indicates that the model is not overfitting and can generalize well. Fig.~\ref{fig:SSD_min_max} compares detections with the maximum and minimum average IoU values under the condition that the confidence level threshold equals 0.5 within the images where both patterns were detected. The model is pretty successful in detecting the patterns covered by ground truth annotations both for whole validation data and observational images in the validation dataset, as seen in Fig.~\ref{fig:SSD_cfm}. The figure presents the confusion matrix, which serves as a key tool for assessing model performance by comparing true labels to predicted labels. It displays the percentage of true positives (correct predictions of the positive class), true negatives (correct predictions of the negative class), false positives (incorrect predictions of the positive class), and false negatives (incorrect predictions of the negative class). This matrix enables us to assess overall accuracy while also highlighting the model's performance for each specific class. As seen from the figure, the percentages of the true positives and true negatives are satisfactorily high while 0.36\% and 0.26\% of the observational dataset identified as background. The 'background' refers to regions of the image that do not contain any objects of interest, specifically areas that are not annotated with bounding boxes for the target classes in the dataset. When evaluating the model's performance using a confusion matrix, the background is crucial for determining false positives (incorrectly identified objects in the background), in addition to other performance parameters. PASCAL \citep{eve10} and COCO \citep{lin14} evaluation metrics for the model were calculated using the validation data and listed in Table~\ref{tab:metrics}. The Precision-Recall curves for different IoU thresholds are shown in Fig.~\ref{fig:SSD_pr} which indicates an expected decrement in recall for increasing precision. In this context, the precision measures the proportion of true positives among all predicted positives, while recall measures the proportion of true positives detected out of all actual positives. An ideal precision-recall curve for an object detection model should have a high area under the curve, with both precision and recall values remaining consistently high across different decision thresholds, such as Intersection over Union (IoU) values. In a perfect model, precision would remain high even as recall increases, indicating that the model detects most objects without generating false positives. However, in real-world models, precision typically decreases as recall increases. The goal is to achieve a curve that stays high, reflecting strong precision even when recall is high, which indicates a good balance between detecting true positives and avoiding false positives. This balance is critical in object detection tasks for maintaining accurate predictions. \citet{eve10} provided an overview of precision and recall evaluation metrics in object detection tasks, especially within the framework of Pascal VOC, which have been widely used to benchmark object detection models. In the figure, the final precision also decreases in higher IoU values. The minimum patterns seem easy to be detected compared to the pulsation patterns which show more diversity in their shapes. 

\footnotetext{{\url{https://colab.research.google.com/drive/1ndiIPMKumhc8PO48S-b-hgwPfpdMQgz3?usp=sharing}}\label{fn:loss_map}}

\begin{table*}
    \caption{PASCAL and COCO evaluation metrics from detection on the validation set using different models. AP, AR and mAP stand for Average Precision, Average Recall and mean Average Precision. The IoU thresholds are written in parentheses for APs where IoU=0.50:0.05:0.95 refer IoU interval between 0.50 and 0.95 with 0.05 increment. AR (max=1) corresponds to one detection per image value. NPCNN stands for the custom non-pretrained CNN-based model.}
    \label{tab:metrics}
    \centering
    \begin{tabular}{lccccc}
        \hline
         Metric   & SSD &Faster~R-CNN&YOLO&EfficientDet~D1&NPCNN\\
        \hline    
        PASCAL&\\
        \cline{1-1}
        AP$_{P}$ (IoU=0.50) & 0.992& 0.991&0.989 &0.965&0.409\\
        AP$_{M}$ (IoU=0.50) & 0.959& 0.998&0.999 &0.895&0.201\\
        mAP (IoU=0.50) & 0.975&0.994&0.993 &0.929&0.305\\
        \hline
        COCO&\\
        \cline{1-1}
        AP (IoU=0.50:0.05:0.95) & 0.839&0.833 &0.915 & 0.726&0.086\\
        AP (IoU=0.50) & 0.969&0.990 & 0.985&0.925&0.298\\
        AP (IoU=0.75) & 0.952&0.969 &0.964 &0.868&0.028\\
        AR (max=1) & 0.855 &0.848& 0.919&0.762&0.202\\
        \hline
    \end{tabular}
\end{table*}

\subsection{Faster R-CNN Model}\label{sec:frcnn}

Faster Region-based Convolutional Neural Network \citep[Faster R-CNN,][]{ren17} is another effective model for object detection tasks in computer vision. It builds based on its predecessors, R-CNN \citep{ric14} and Faster R-CNN \citep{gir15}, by using a Region Proposal Network \citep[RPN,][]{ren15}  that significantly speeds up object detection. 

\begin{figure*}
   \centering
   \includegraphics[width=0.8\textwidth]{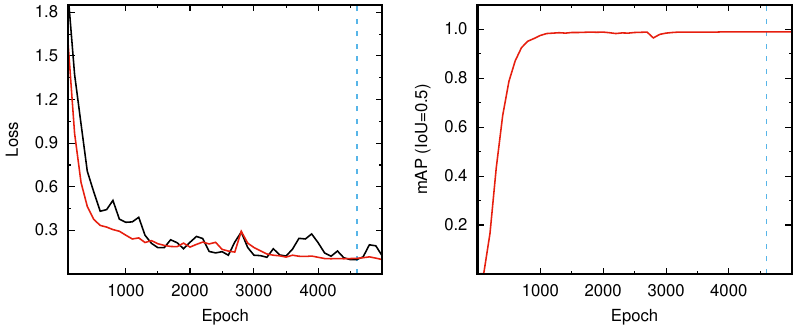}
   \caption{The variation of Loss and mAP across epochs for the Faster R-CNN model is shown. The black and red lines represent training and validation values, respectively, while the blue dashed line indicates the epoch corresponding to the final model. TensorBoard parameter plots are available in the accompanying Google Colab notebook\footref{fn:loss_map}.}
   \label{fig:frcnn_loss_map}%
    \end{figure*}

\begin{figure*}
\centering
\begin{tabular}{ |c|c| }
\hline
\includegraphics[width=0.35\textwidth]{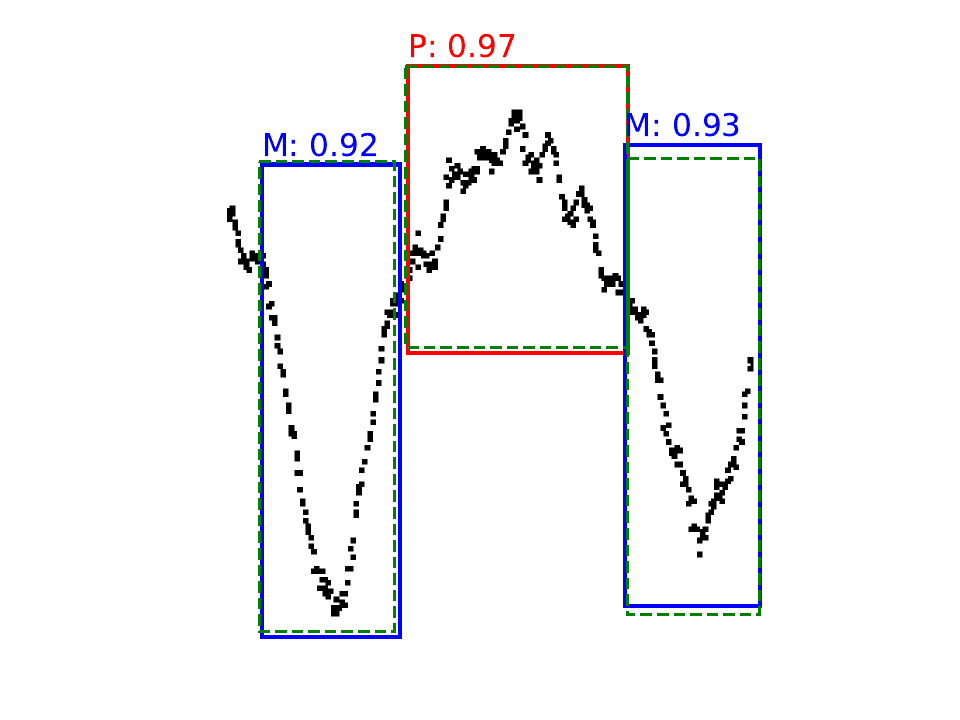} & \includegraphics[width=0.35\textwidth]{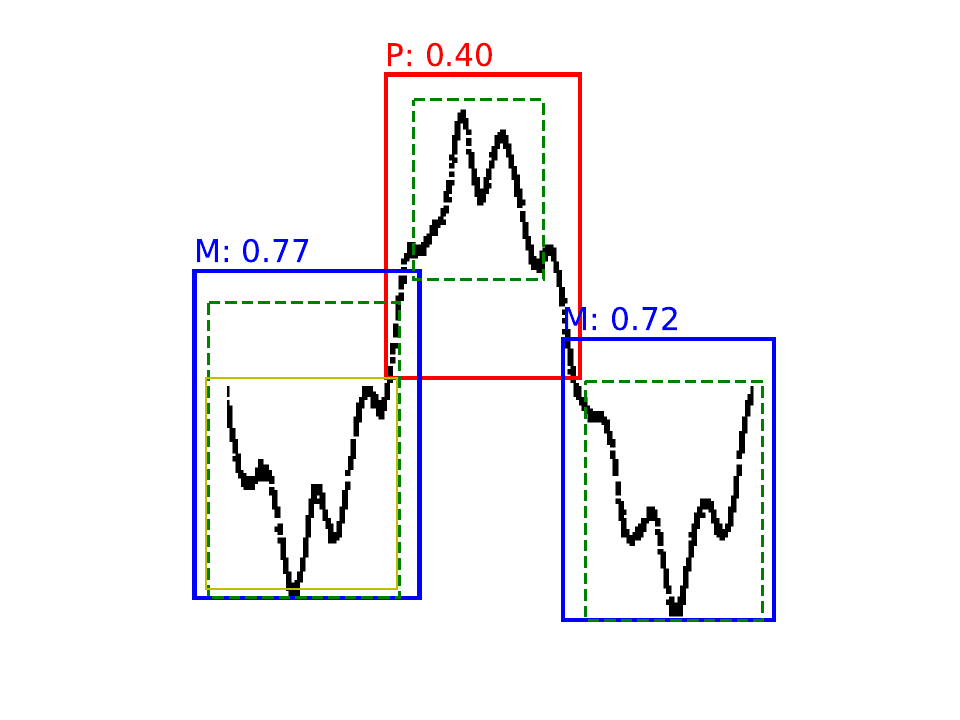} \\
\hline
\end{tabular}
\caption{Examples of varying detection performance of Faster R-CNN model on observational light curve images from the validation set, displaying predicted bounding boxes for each ground truth annotation. Detected patterns (in red and blue bounding boxes) with the highest (left, TIC~158794976) and lowest (right, TIC~232637376) average IoU values are compared against ground truth annotations (green dashed boxes). The numbers above each box indicate the IoU value for the corresponding detection. "P" and "M" denote pulsation and minimum patterns, respectively. The yellow box on the right panel represents the additional detection that is not present in the ground truth annotations.}
 \label{fig:frcnn_min_max}
\end{figure*}

\begin{figure*}
\centering
\begin{tabular}{ |c|c| }
\hline
\begin{minipage}{0.35\textwidth}
            \centering
           \includegraphics[width=0.9\linewidth]{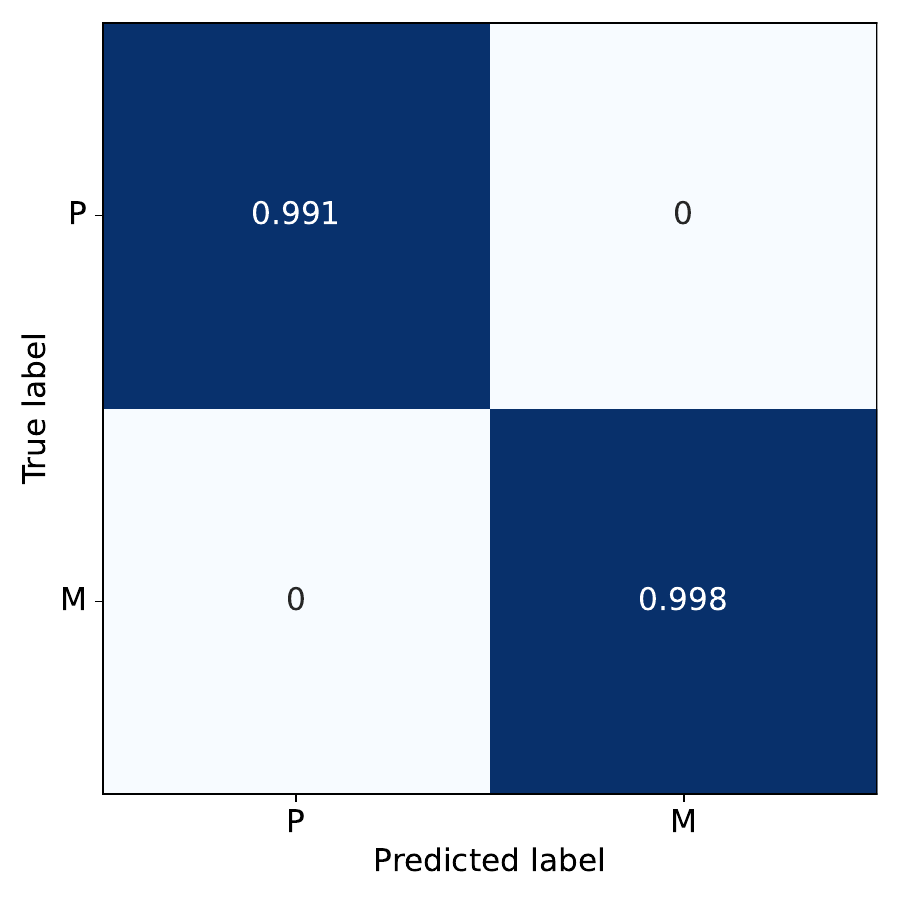}
        \end{minipage}
        &
        \begin{minipage}{0.35\textwidth}
            \centering
            \includegraphics[width=0.9\linewidth]{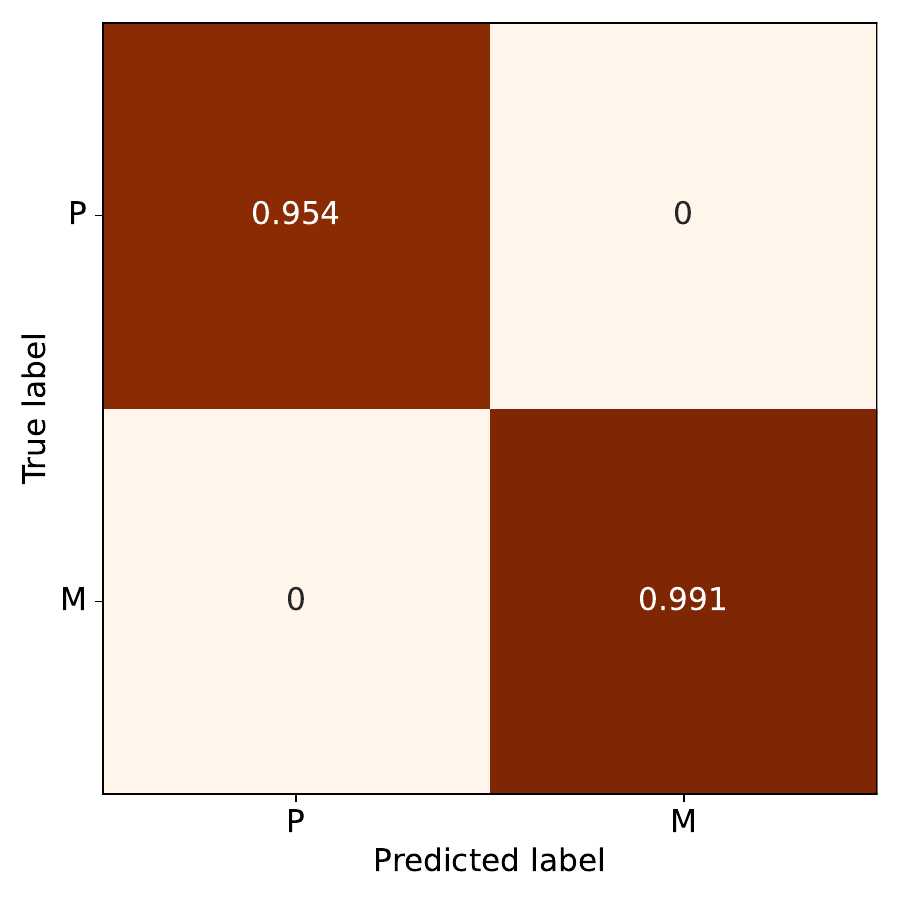}
        \end{minipage} \\

\hline
\end{tabular}
\caption{The confusion matrices for the overall data (left) and observational data only (right) in the validation dataset are calculated using predicted class and bounding boxes, with the IoU and confidence threshold set to 0.5 with the Faster R-CNN model. "P" and "M" denote pulsation and minimum patterns, respectively.}
 \label{fig:frcnn_cfm}
\end{figure*}

\begin{figure*}
\centering
\includegraphics[width=\textwidth]{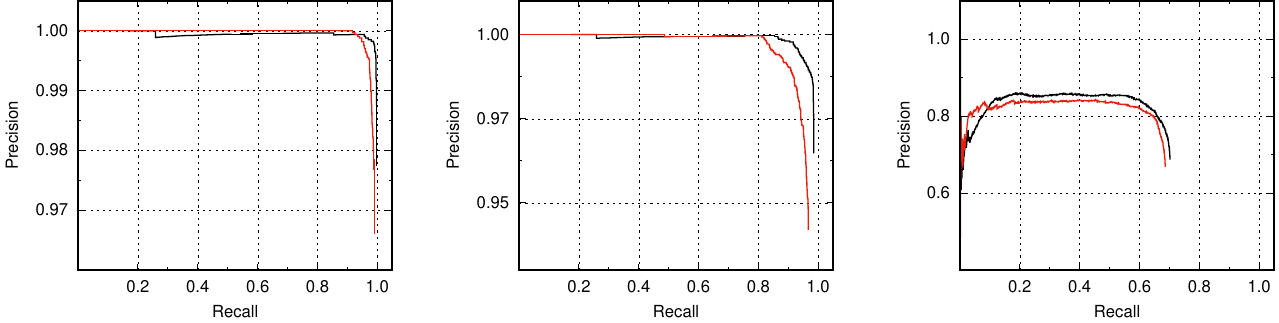} 
\caption{Precision-Recall curves for the Faster R-CNN model at IoU thresholds of 0.5 (left), 0.75 (middle), and 0.9 (right) are shown. The black and red lines represent minimum and pulsation patterns, respectively.}
 \label{fig:frcnn_pr}
\end{figure*}

Training until the final epoch, 4600th, lasted 25 min 28 sec on one Nvidia GeForce RTX 2080 Ti accelerator. The decrement of the total loss function, which is the sum of localization, classification and objectness losses, during training and evaluation is demonstrated along with the mean average precision in Fig.~\ref{fig:frcnn_loss_map}. Even the selection of small batches shows itself with fluctuations in the curve, the decrement is smooth, and further critical improvement was not observed in the following epochs. The validation loss value is close to the training at the 4600th epoch, which corresponds to the final model's ability to generalize the unseen data. Detections were conducted using the final model file\footref{fn:models}, which takes 0.07$\,\mathrm{seconds}$ per frame, and the samples of inferences with the highest and lowest average IoU values are shown in Fig.~\ref{fig:frcnn_min_max}.  The model can be considered to detect patterns where pulsations are superposed on the minimum pattern (right panel) similar to SSD, although the IoU values are around 0.7. PASCAL and COCO metrics of the detection are presented in Table~\ref{tab:metrics}. The percentage of correctly predicted detections are shown in Fig.~\ref{fig:frcnn_cfm} for the complete dataset and observations, separately, which are above 99\% for three classes. The precision-recall curves (Fig.~\ref{fig:frcnn_pr}), on the other hand, indicate that the model performance decreases with increasing IoU threshold, as expected.

\subsection{YOLO Model}
YOLO \citep[You Only Look Once,][]{red16} uses a unique methodology other than two-step methods, which is conventionally shown in regional scanning techniques. YOLO runs based on dividing the input image into parts and estimates the bounding boxes and class probabilities for a detected pattern within the individual parts. The advantage of the methodology of the YOLO lies in its speed and accuracy. The algorithm has wide usage areas in various disciplines.

\begin{figure*}
   \centering
   \includegraphics[width=0.8\textwidth]{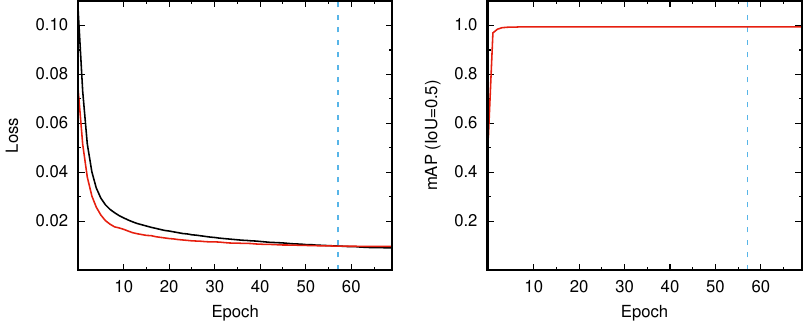}
   \caption{The plot shows Loss and mAP as they change with increasing epochs for the YOLO model. The black and red lines represent training and validation values, respectively, and the blue dashed line marks the epoch of the final model. TensorBoard parameter plots are available in the provided Google Colab notebook\footref{fn:loss_map}.}
   \label{fig:yolo_loss_map}%
    \end{figure*}

\begin{figure*}
\centering
\begin{tabular}{ |c|c| }
\hline
\includegraphics[width=0.35\textwidth]{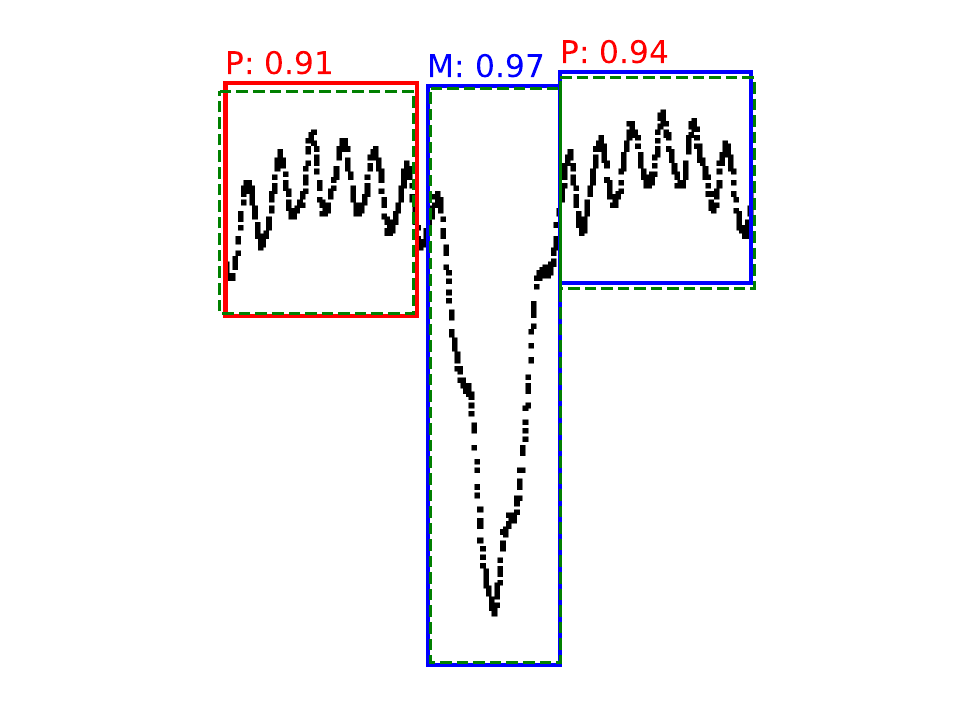} & \includegraphics[width=0.35\textwidth]{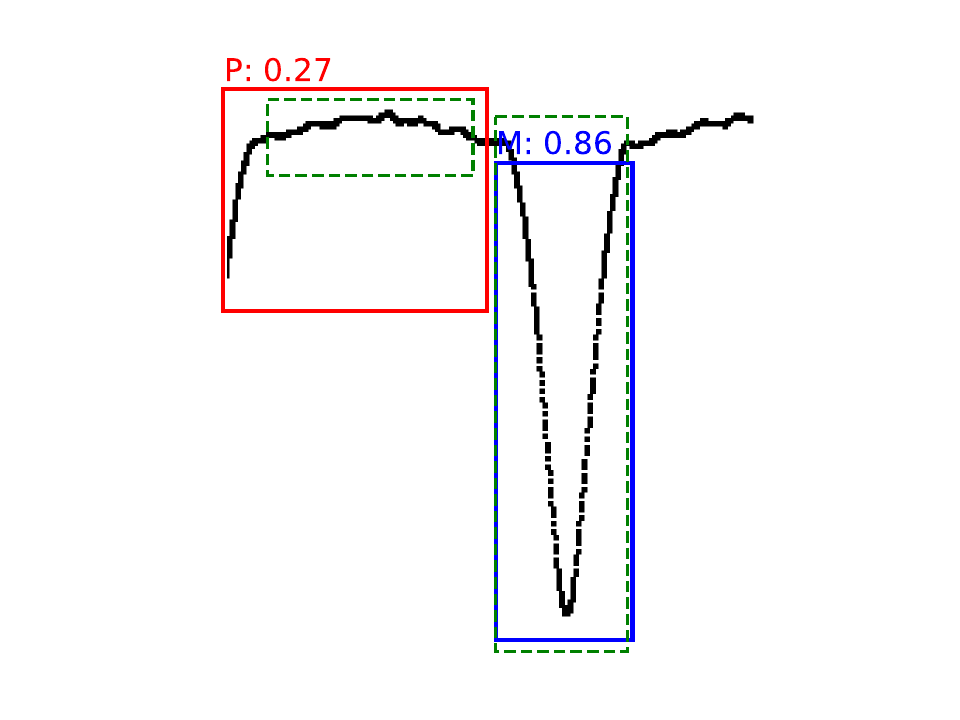} \\
\hline
\end{tabular}
\caption{Different detection performances on observational light curve images from the validation set, showing predicted bounding boxes for each ground truth annotation using YOLO model. The detected patterns, enclosed in red and blue boxes, with the highest (TIC~354922610, left) and lowest (TIC~323292655, right) average IoU values, are compared to the ground truth annotations marked by green dashed boxes. The numbers above each box indicate the IoU value for that detection. "P" and "M" represent pulsation and minimum patterns, respectively.}
 \label{fig:yolo_min_max}
\end{figure*}

\begin{figure*}
\centering
\begin{tabular}{ |c|c| }
\hline
\begin{minipage}{0.35\textwidth}
            \centering
           \includegraphics[width=0.9\linewidth]{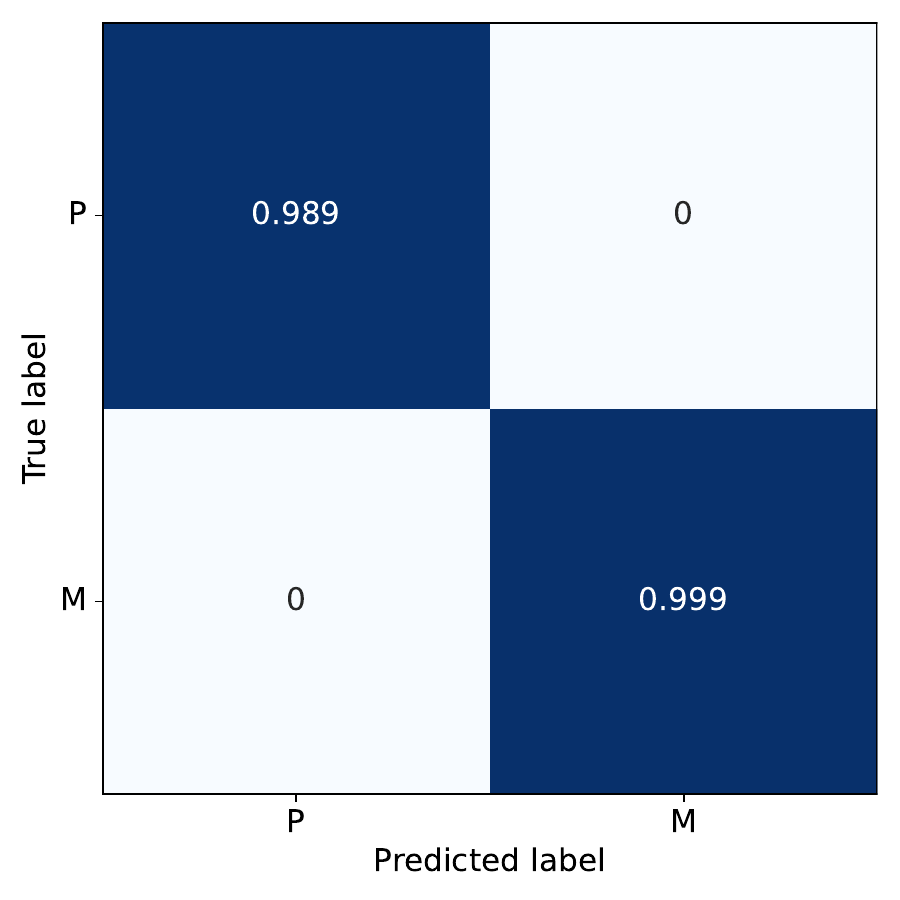}
        \end{minipage}
        &
        \begin{minipage}{0.35\textwidth}
            \centering
            \includegraphics[width=0.9\linewidth]{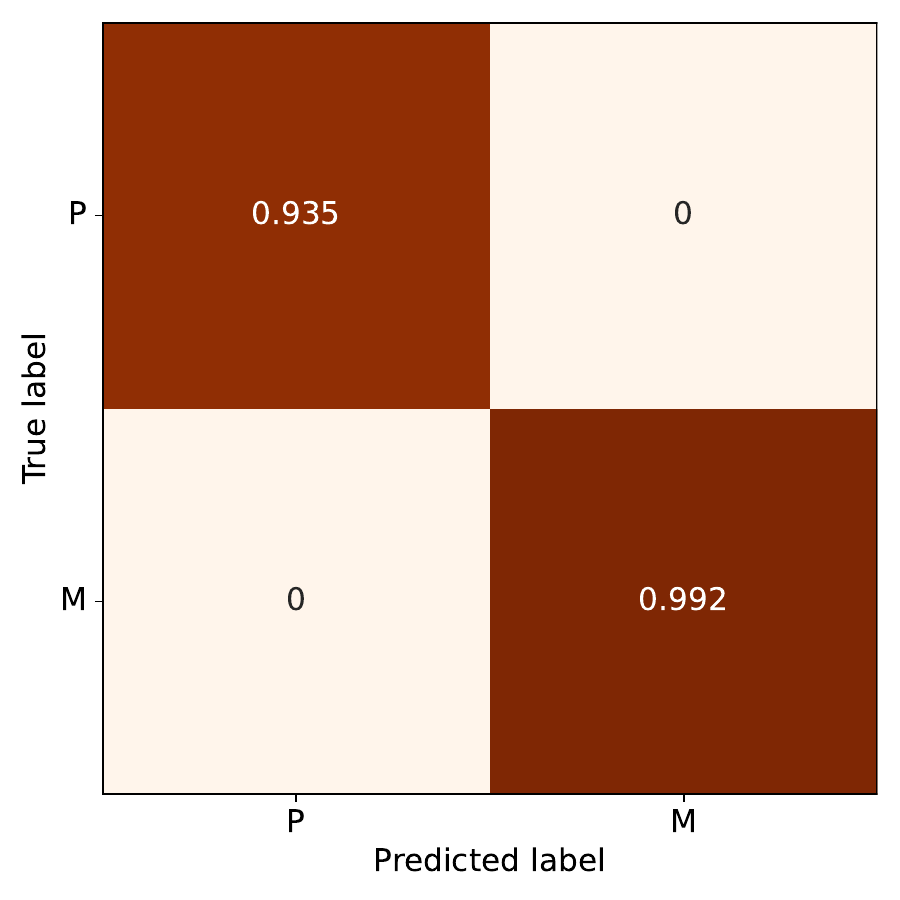}
        \end{minipage} \\
\hline
\end{tabular}
\caption{The confusion matrices for the overall data (left) and observational data (right) in the validation dataset are derived based on predicted classes and bounding boxes using YOLO model, with IoU and confidence threshold set at 0.5. "P" and "M" stand for pulsation and minimum patterns.}
 \label{fig:yolo_cfm}
\end{figure*}

\begin{figure*}
\centering
\includegraphics[width=\textwidth]{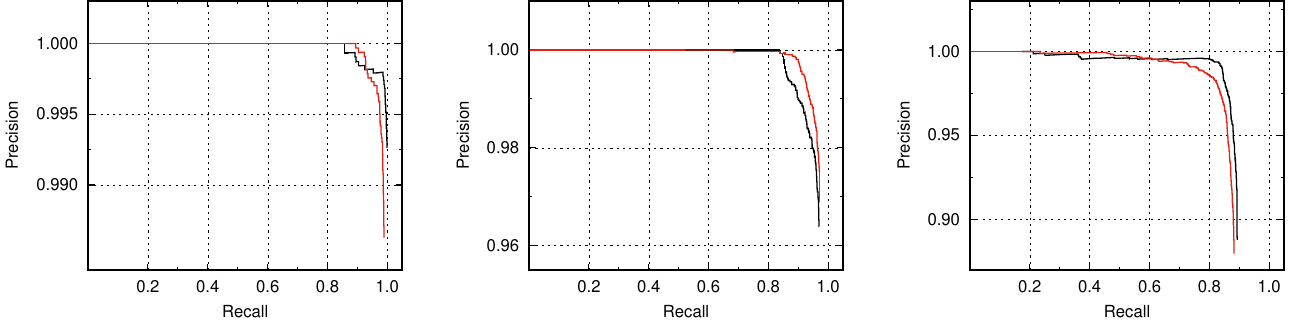} 
\caption{The YOLO model’s Precision-Recall curves are illustrated for IoU thresholds of 0.5 (left), 0.75 (middle), and 0.9 (right). Minimum patterns are indicated by the black line, while pulsation patterns are indicated by the red line.}
 \label{fig:yolo_pr}
\end{figure*}

During the training process, we employed YOLOv5s architecture together with the pre-trained weights, a compact version within the YOLOv5\footnote{\url{https://github.com/ultralytics/yolov5}\label{fn:yol}} family, designed to offer faster inference times while maintaining reasonable accuracy. The training and validation losses decreased very smoothly by increasing steps (Fig.~\ref{fig:yolo_loss_map}) and no improvements were observed after the 57th epoch, the step where the final model\footref{fn:models} extracted. The loss function in the figure is the total loss, the summation of three losses, box loss, class loss and objectness loss. The training process lasted 5 $\,\mathrm{hours}$ and 43 $\,\mathrm{minutes}$, substantially longer compared to other models. The model can be considered to show a very good performance in detection patterns in the validation dataset, which can be seen in Fig.~\ref{fig:yolo_min_max}, where the detections with maximum and minimum average IoU values are shown. Contrary to the long training process, the detection over 3411 images took 33.92$\,\mathrm{seconds}$, averaging approximately 0.01$\,\mathrm{seconds}$ per image, the best performance among all our models. The accurate detection of the model, even for the observational data, can be seen in the confusion matrices in Fig.~\ref{fig:yolo_cfm} where the results are close to that of Faster R-CNN. The precision-recall curves of the model for different IoU values show consistent variation compared to all other models in the study, as given in Fig.~\ref{fig:yolo_pr}. The shapes underscore the model's robustness in balancing precision and recall effectively, even under different intersection-over-union settings as we remarked in Sec.~\ref{sec:ssd}. The resulting metrics of the detection are listed in Table~\ref{tab:metrics}.

\subsection{EfficientDet~D1 Model}
EfficientDet \citep{tan20}, combines various techniques to achieve high accuracy and efficiency simultaneously. The balance between accuracy and computational efficiency complies with a wise usage of the scaling method, optimizing model architecture, resolution, and input size.

\begin{figure*}
   \centering
   \includegraphics[width=0.8\textwidth]{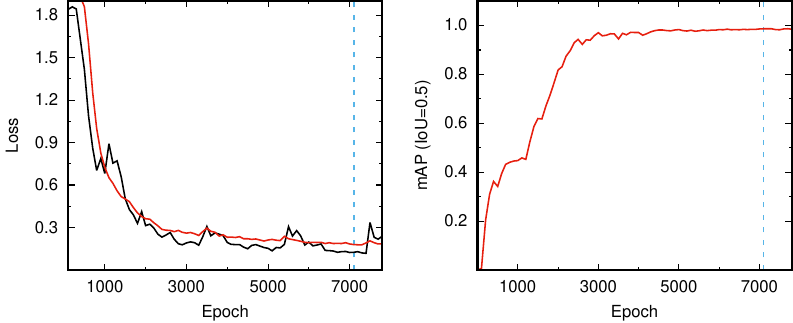}
   \caption{The decrement in Loss and mAP with increasing epochs for the EfficientDet D1 model is illustrated. The black and red lines denote training and validation values, respectively. The blue dashed line marks the final model at the 7100th epoch. TensorBoard parameter plots are available in the provided Google Colab notebook\footref{fn:loss_map}.}
   \label{fig:Eff_loss_map}%
    \end{figure*}

\begin{figure*}
\centering
\begin{tabular}{ |c|c| }
\hline
\includegraphics[width=0.35\textwidth]{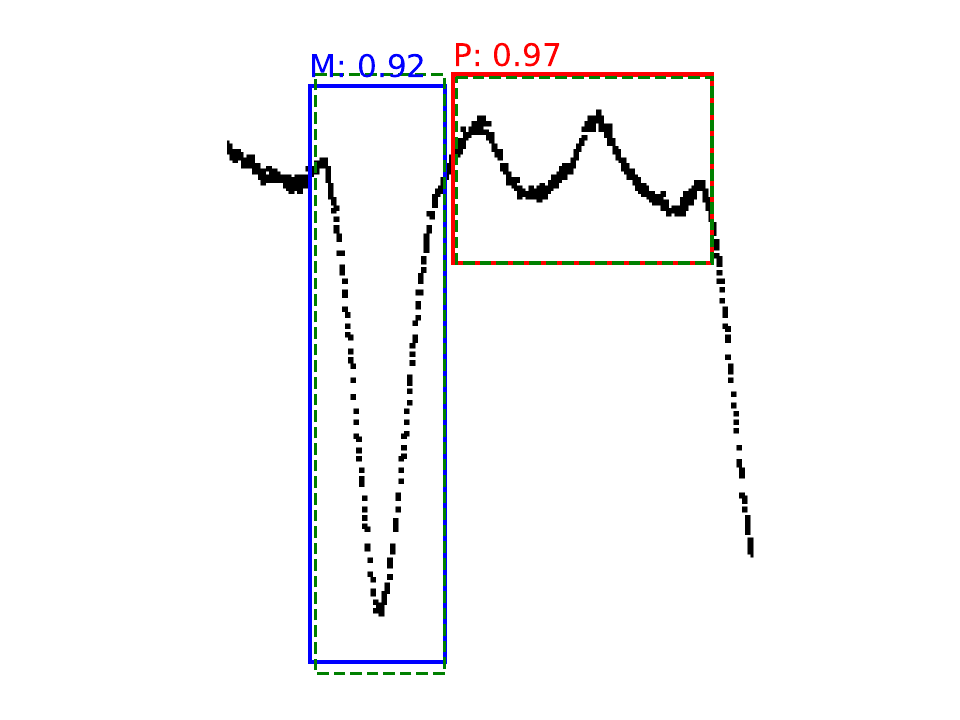} & \includegraphics[width=0.35\textwidth]{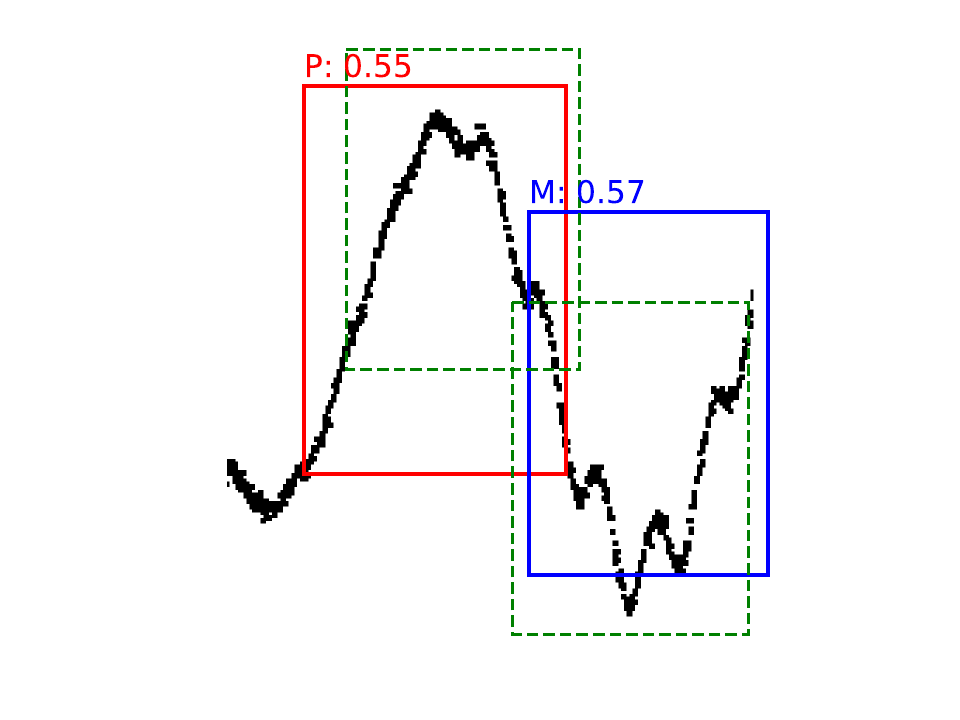} \\
\hline
\end{tabular}
\caption{Examples of detection performance variation of the EfficientDet~D1 model on observational light curve images from the validation set, featuring predicted bounding boxes for each ground truth annotation. Detected patterns, highlighted in red and blue boxes, with the highest (left, TIC~350030939) and lowest (right, TIC~355151781) average IoU values are compared to the ground truth annotations shown in green dashed boxes. The numbers above the boxes represent the IoU values for each detection. "P" and "M" stand for pulsation and minimum patterns, respectively.}
 \label{fig:Eff_min_max}
\end{figure*}

\begin{figure*}
\centering
\begin{tabular}{ |c|c| }
\hline
\begin{minipage}{0.35\textwidth}
            \centering
           \includegraphics[width=0.9\linewidth]{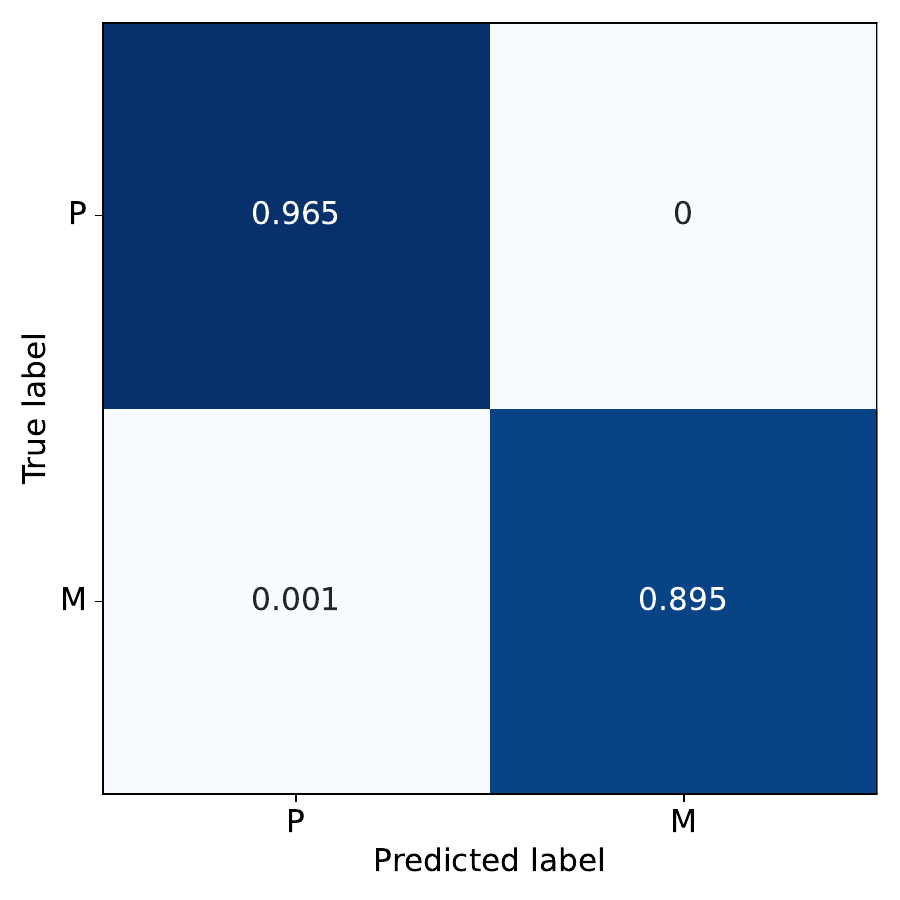}
        \end{minipage}
        &
        \begin{minipage}{0.35\textwidth}
            \centering
            \includegraphics[width=0.9\linewidth]{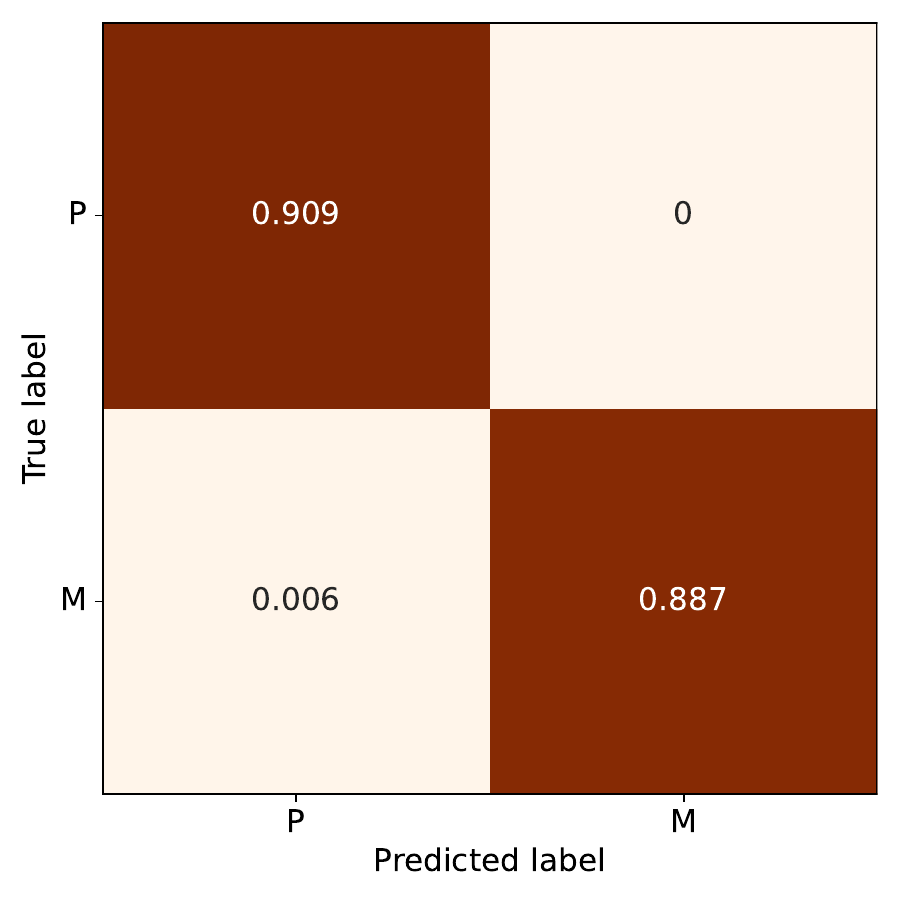}
        \end{minipage} \\

\hline
\end{tabular}
\caption{The confusion matrices for the overall dataset (left) and only observational data (right) in the validation set are generated based on predicted classes and bounding boxes using the EfficientDet~D1 model and setting an IoU and confidence threshold of 0.5. Consider that 0.1\% and 0.6\% of the minimum patterns are misclassified as pulsation for all (left) and observational data (right), respectively. "P" and "M" denote pulsation and minimum patterns.}
 \label{fig:Eff_cfm}
\end{figure*}

\begin{figure*}
\centering
\includegraphics[width=\textwidth]{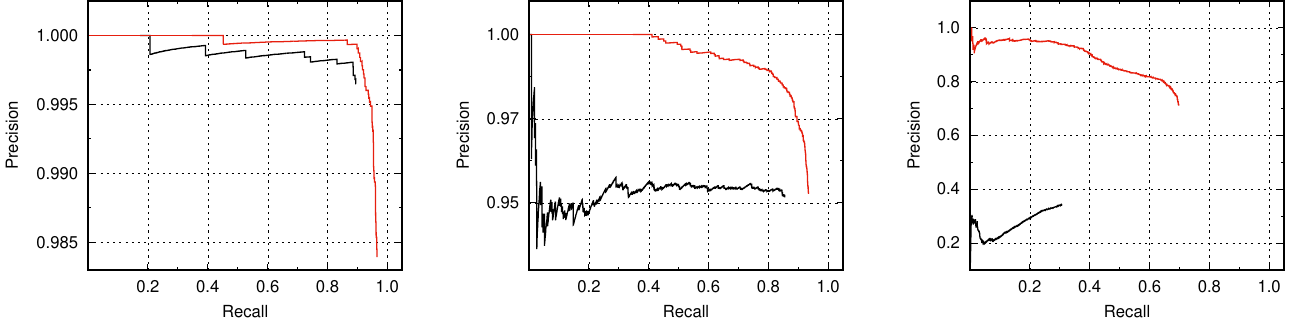} 
\caption{Precision-Recall curves for the EfficientDet~D1 model at IoU thresholds of 0.5 (left), 0.75 (middle), and 0.9 (right). The black line corresponds to minimum patterns and the red line to pulsation patterns.}
 \label{fig:Eff_pr}
\end{figure*}

We applied two consecutive training procedures similar to that mentioned in Sec.~\ref{sec:ssd}. The retraining process lasted 23$\,\mathrm{minutes}$ 25$\,\mathrm{seconds}$ corresponding to 7100 epochs. The detection per frame took 0.58$\,\mathrm{seconds}$ using the final model file\footref{fn:models}, the longest time of any model. The training and validation loss through the epoch number is shown in Fig.~\ref{fig:Eff_loss_map} with the mAP values. The prominent fluctuations in the training curve rise from the small batch size mentioned previously. The loss values beyond the 7100th epoch tend not to improve in the following epochs. mAP(Iou=50), on the other hand, exceeds 0.9. The detections with maximum and minimum average IoUs are plotted in Fig.~\ref{fig:Eff_min_max} where values around 0.5 can be seen. The confusion matrices are given in Fig.~\ref{fig:Eff_cfm} while the related metrics are listed in Table~\ref{tab:metrics}. The model performed relatively weakly in detecting all the minimum patterns, whereas it successfully predicts more than 90\%  of the pulsation patterns in observational data. The precision-recall curves for various IoUs are also presented in Fig.~\ref{fig:Eff_pr}. An irregular trend and a large amount of decrement in precision for the minimum pattern with the increasing IoU threshold is clear. It is worth emphasizing that we tried the previous lighter member, EfficientDet D0, however, the results, especially the decrement in the cost function, were not as satisfactory as expected from an EfficientDet family model. 

\subsection{A non-pretrained CNN-based Model}\label{sec:npcnn}

We implemented a CNN-based object detection model from scratch which is not based on a pre-trained architecture to compare the model performance to pretrained ones and to see if our task can be carried out faster and more effectively with a simpler model consisting of CNNs. The model\footref{fn:repo} algorithm was set to be able to read two classes per image, therefore, our annotation files\footnote{\url{https://drive.google.com/file/d/1QazbeRRfrhe5RX8hEbuUZxbcoHHkGit2/view?usp=sharing}} include ground truths of two patterns for each image.

\begin{figure*}
   \centering
   \includegraphics[width=0.8\textwidth]{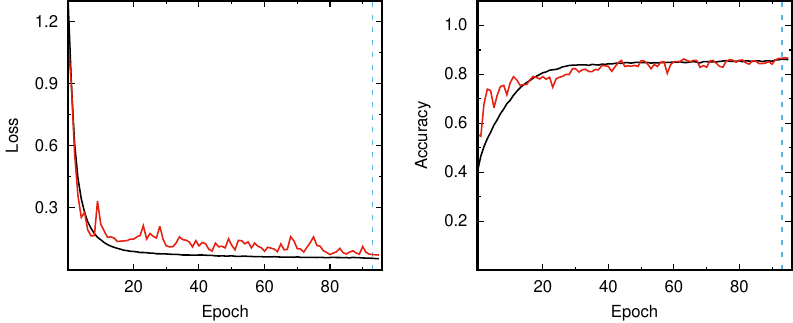}
   \caption{Left: Variation of Loss and mAP for the non-pretrained CNN model. Right: Accuracy of the box predictor across epochs. The black and red lines represent training and validation values, respectively, with the blue dashed line marking the epoch of the final model.}
   \label{fig:cnn_loss_map}%
    \end{figure*}

\begin{figure*}
\centering
\begin{tabular}{ |c|c| }
\hline
\includegraphics[width=0.35\textwidth]{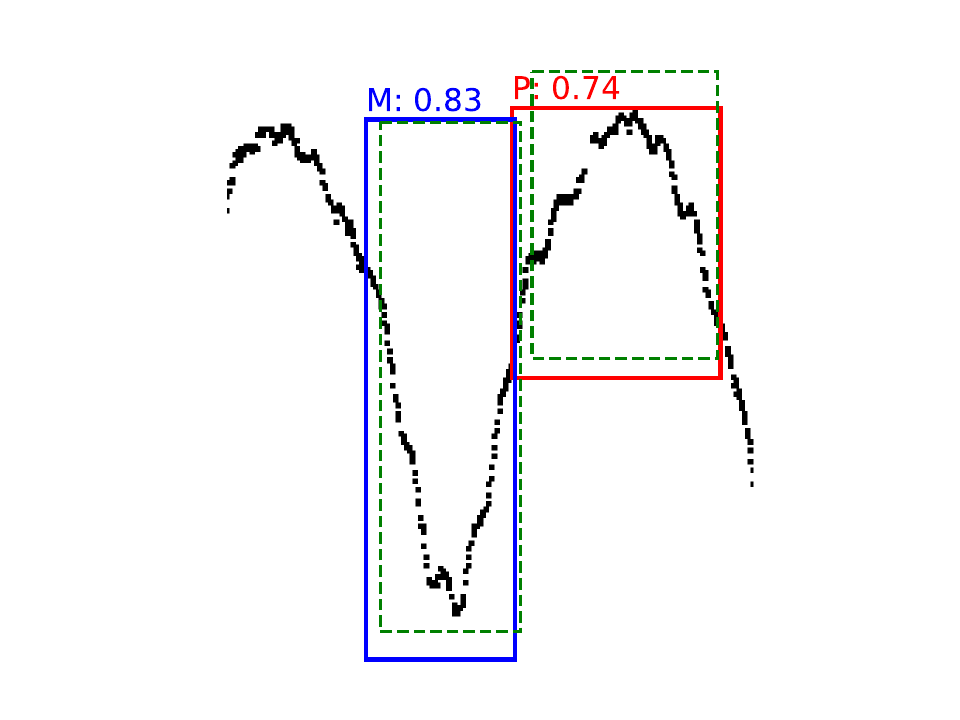} & \includegraphics[width=0.35\textwidth]{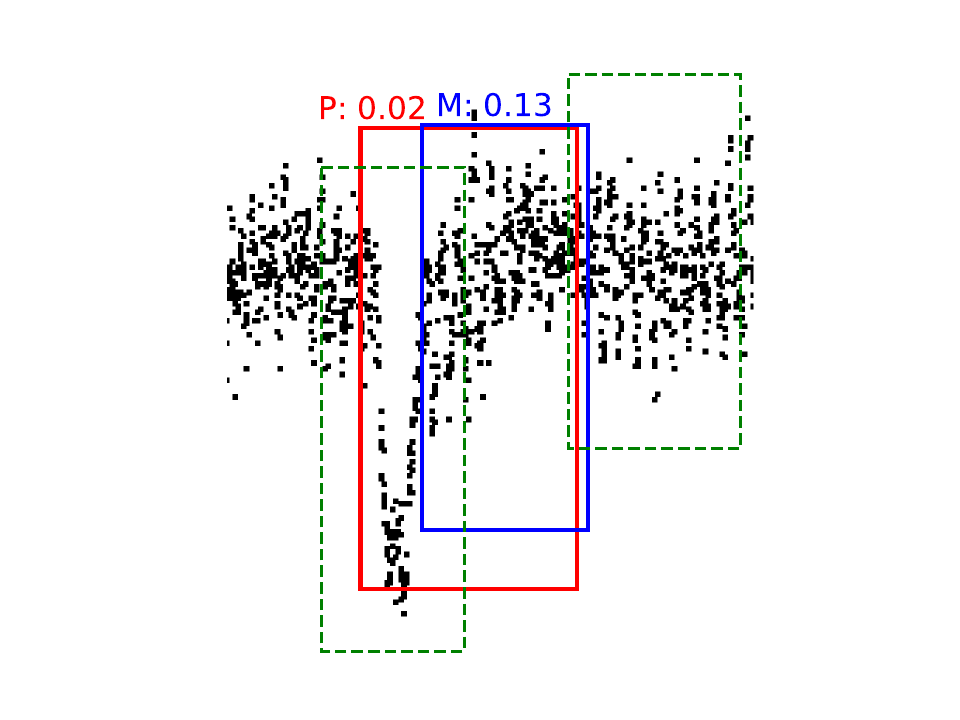} \\
\hline
\end{tabular}
\caption{Samples of detection performance on observational light curve images from the validation set, displaying predicted bounding boxes for each ground truth annotation for the non-pretrained CNN model. Detected patterns, shown in red and blue boxes, with the highest (left, TIC~298734307) and lowest (right, TIC~409934330) average IoU values are compared to the ground truth annotations marked by green dashed boxes. The numbers above each box represent the IoU value for the respective detection. "P" and "M" denote pulsation and minimum patterns, respectively.}
 \label{fig:cnn_min_max}
\end{figure*}

\begin{figure*}
\centering
\begin{tabular}{ |c|c| }
\hline
\begin{minipage}{0.35\textwidth}
            \centering
           \includegraphics[width=0.9\linewidth]{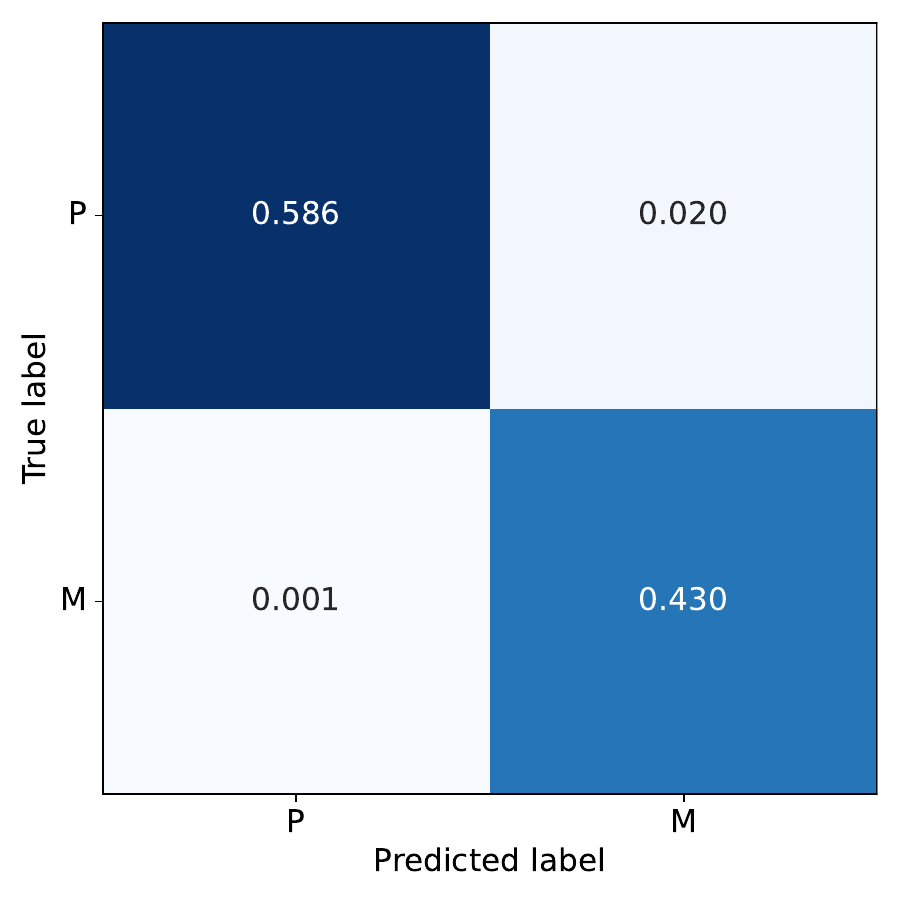}
        \end{minipage}
        &
        \begin{minipage}{0.35\textwidth}
            \centering
            \includegraphics[width=0.9\linewidth]{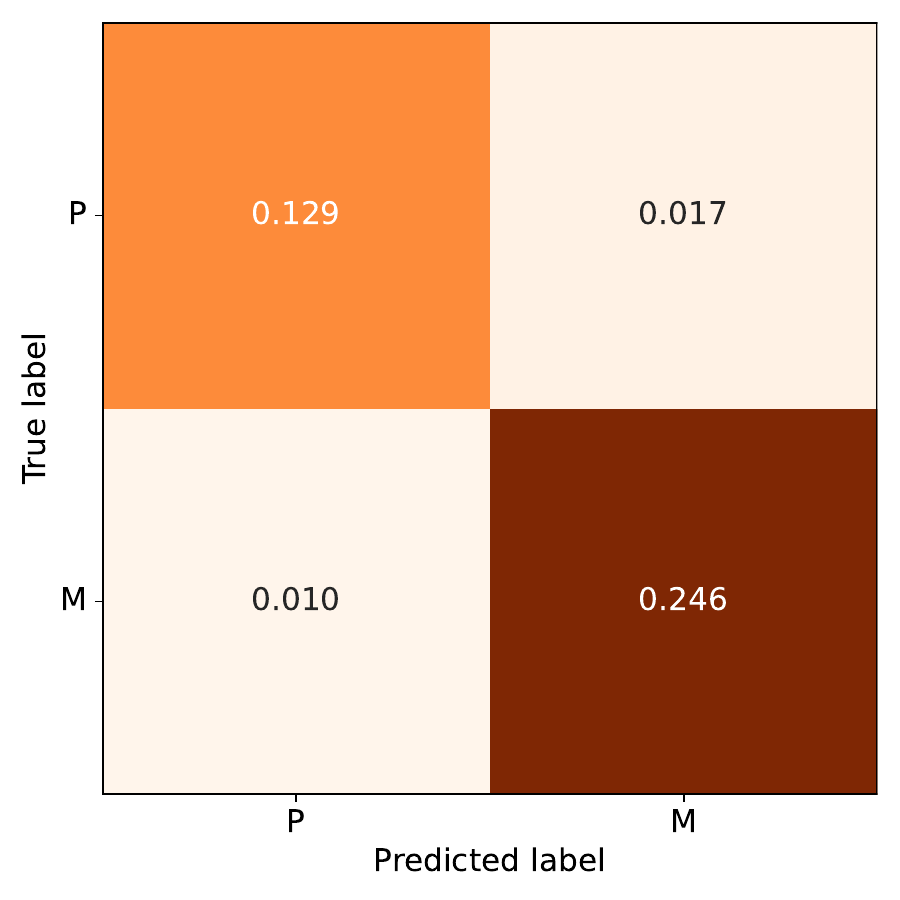}
        \end{minipage} \\
\hline
\end{tabular}
\caption{The confusion matrices for the overall data (left) and observational data (right) in the validation set are computed based on predicted classes and bounding boxes, with IoU and confidence thresholds set to 0.5, using the non-pretrained CNN-based model. We note that the model is considerably unsuccessful in detecting patterns, especially on observational data. "P" and "M" denote pulsation and minimum patterns, respectively.}
 \label{fig:cnn_cfm}
\end{figure*}

\begin{figure*}
\centering
\includegraphics[width=\textwidth]{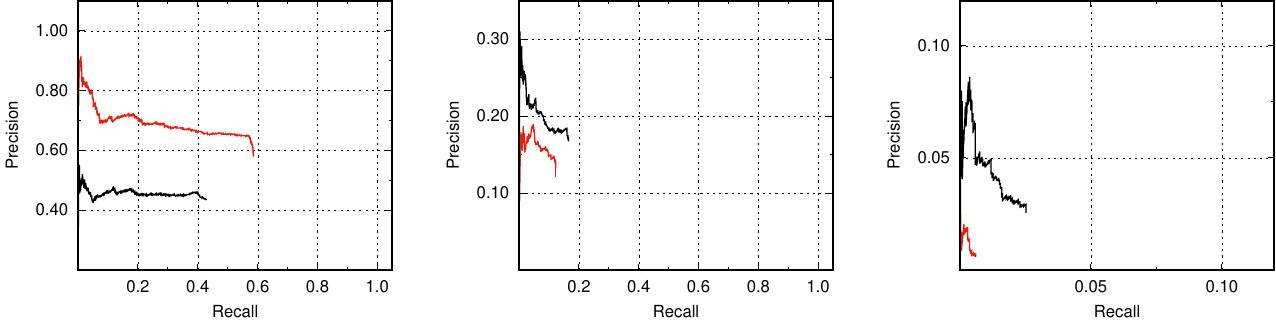} 
\caption{The Precision-Recall curves for the non-pretrained CNN model are plotted with IoU thresholds of 0.5 (left), 0.75 (middle), and 0.9 (right). The black line denotes the minimum, and the red line indicates pulsation patterns. Be aware that the Recall axis is shortened in the rightmost plot for clear visibility.}
 \label{fig:cnn_pr}
\end{figure*}

The training lasts 2 $\,\mathrm{hours}$ 24 $\,\mathrm{minutes}$ on the same accelerator mentioned earlier. The best model\footref{fn:models} was achieved in the 93rd epoch where the train and validation losses are almost at their minimum without significant overfitting. The learning curves are presented in Fig.~\ref{fig:cnn_loss_map} where the loss is the summation of classification and box regression losses. It exhibited a consistent improvement in both loss and accuracy metrics. The accuracy of the box predictor, which achieves a plateau at about 0.9, was also plotted in the figure. A steep decline in both training and validation loss, accompanied by a steady rise in accuracy, indicates effective learning. The model infers 3411 validation data in 315.65$\,\mathrm{seconds}$, corresponding to almost 0.09$\,\mathrm{seconds}$ per frame. The predictions are noticeably weaker than the previous models such that no detections occur with the average IoU value exceeding 0.8. The metrics obtained from the detections are also listed in Table~\ref{tab:metrics}. The detections with maximum and minimum average IoU values are illustrated in Fig.~\ref{fig:cnn_min_max}. The low quality of the detections can also be seen from the confusion matrices (Fig.~\ref{fig:cnn_cfm}) for entire and observational data separately indicating that the model detected a huge amount of the patterns as background. The Precision-Recall curves for three IoU thresholds are also given in Fig.~\ref{fig:cnn_pr}, corresponding that the increasing threshold value dramatically changes the curve to end up with small precision and recall values.

\begin{figure*}
\centering
\includegraphics[width=\textwidth]{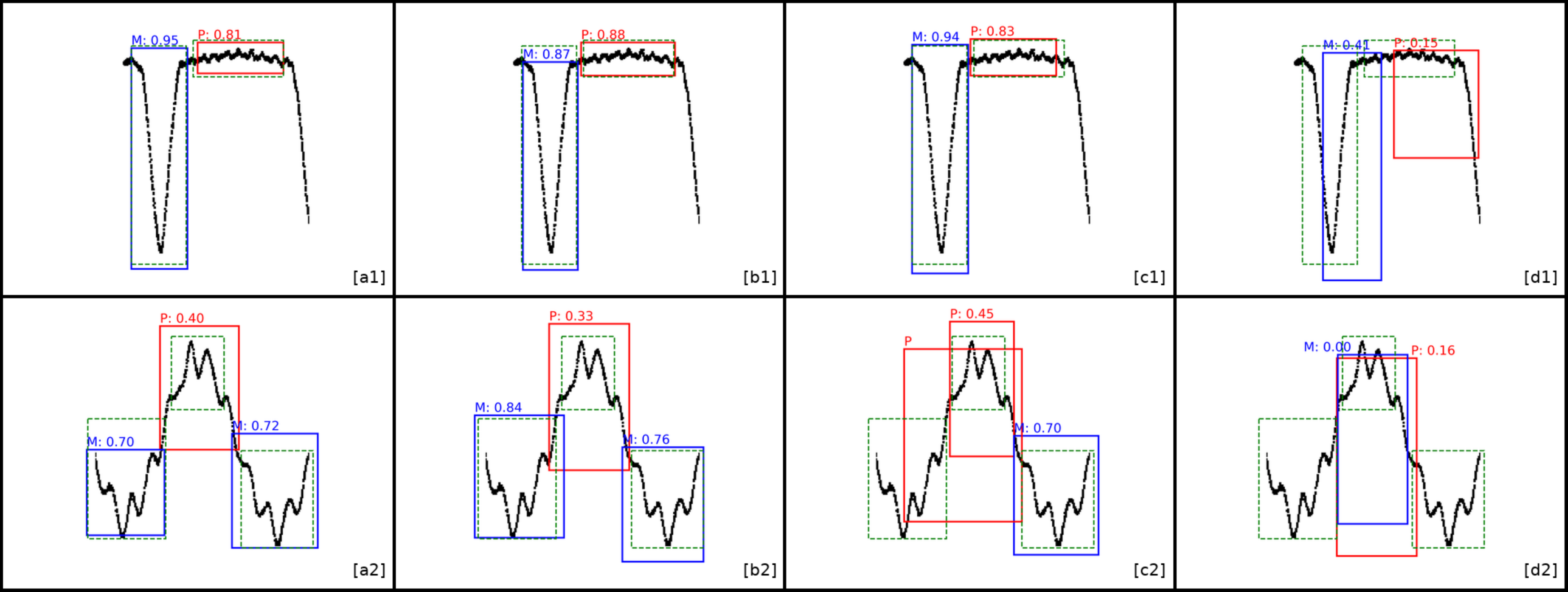} \\
\caption{The detection performance of Faster-RCNN (a1, a2), YOLO (b1, b2),  EfficientDet~D1 (c1,c2)  and non-pretrain CNN-based (d1, d2) models on the light curves in Fig.~\ref{fig:SSD_min_max}.}
 \label{fig:SSD_source}
\end{figure*}

\begin{figure*}
\centering
\includegraphics[width=\textwidth]{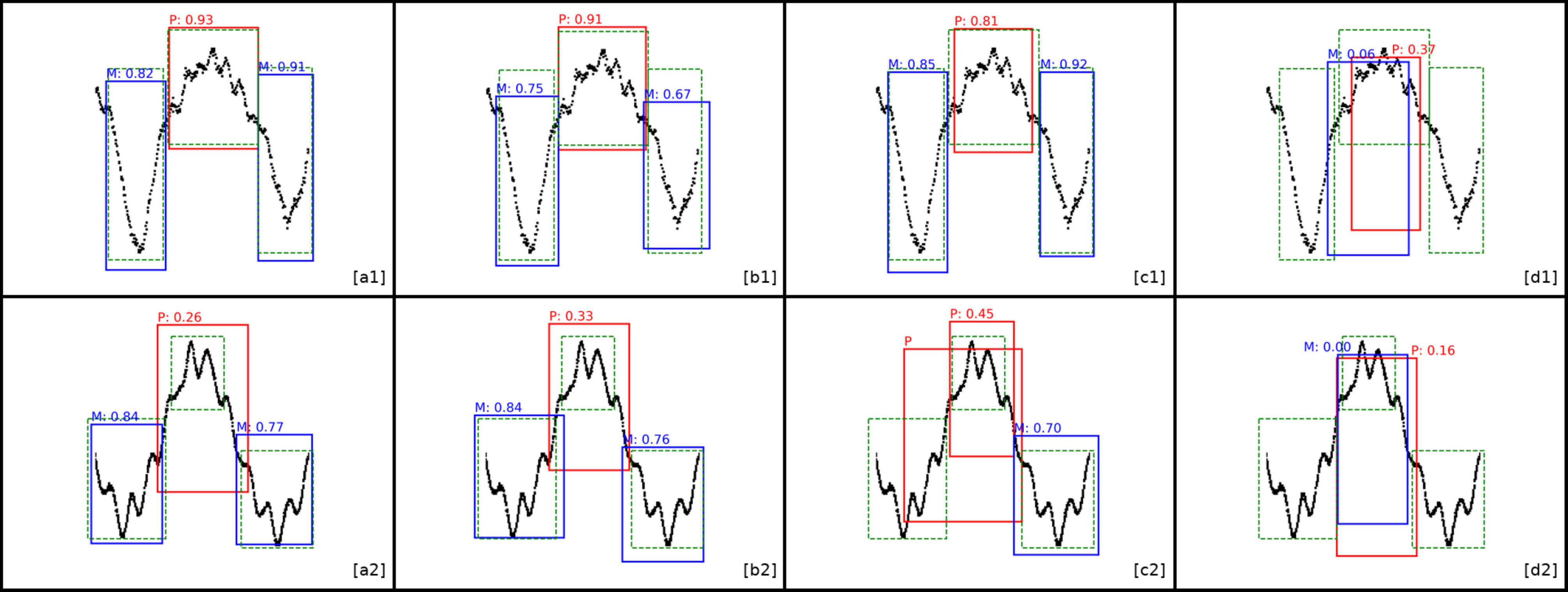} \\
\caption{Same as Fig.~\ref{fig:SSD_source}, but for the detection performance of SSD (a1, a2), YOLO (b1, b2),  EfficientDet~D1 (c1,c2)  and non-pretrain CNN-based (d1, d2) models on the light curves in Fig.~\ref{fig:frcnn_min_max}.}
 \label{fig:frcnn_source}
\end{figure*}

\begin{figure*}
\centering
\includegraphics[width=\textwidth]{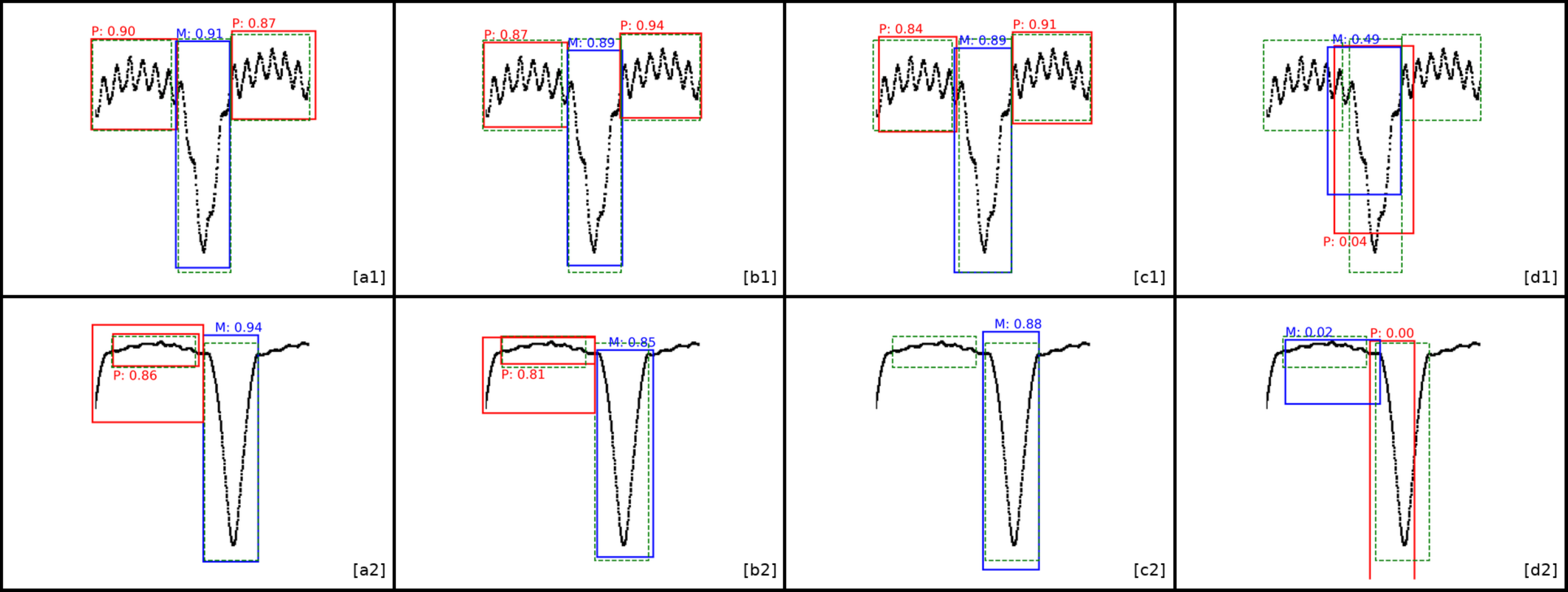} \\
\caption{Same as Fig.~\ref{fig:SSD_source}, but for the detection performance of SSD (a1, a2), Faster-RCNN (b1, b2),  EfficientDet~D1 (c1,c2)  and non-pretrain CNN-based (d1, d2) models on the light curves in Fig.~\ref{fig:yolo_min_max}.}
 \label{fig:yolo_source}
\end{figure*}

\begin{figure*}
\centering
\includegraphics[width=\textwidth]{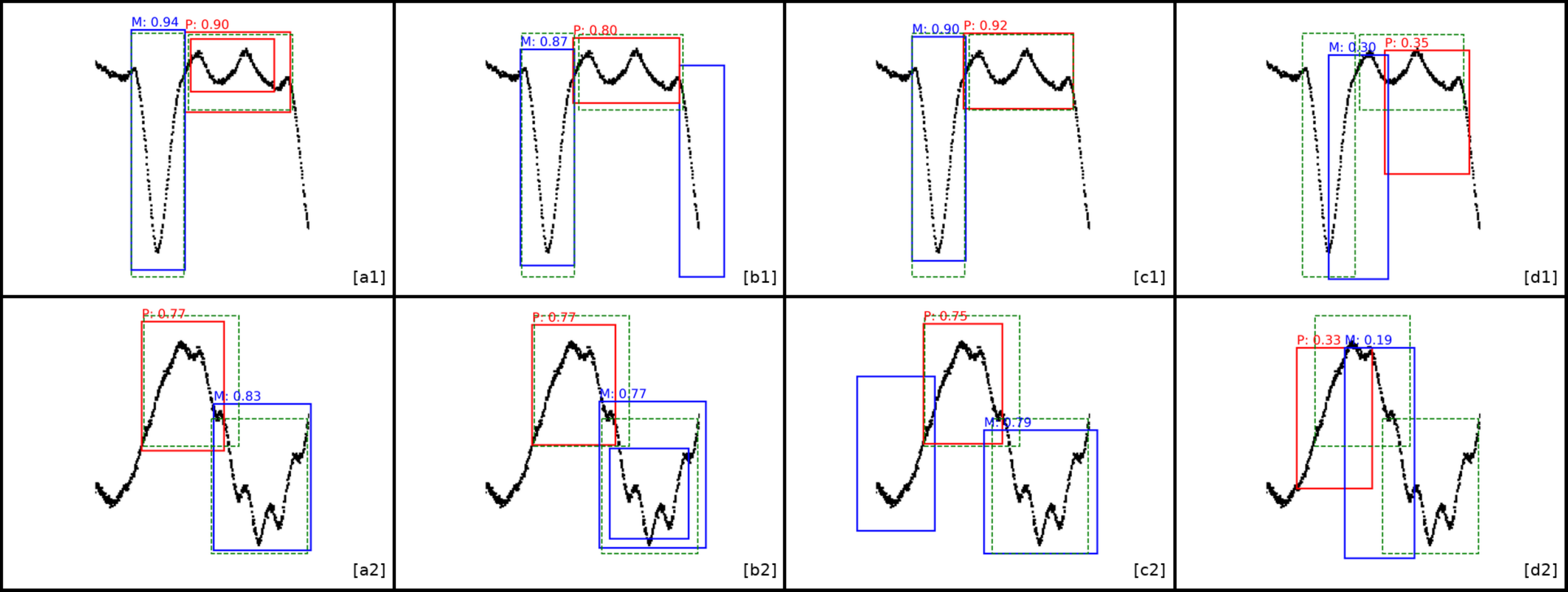} \\
\caption{Same as Fig.~\ref{fig:SSD_source}, but for the detection performance of SSD (a1, a2), Faster-RCNN (b1, b2),  YOLO (c1,c2)  and non-pretrain CNN-based (d1, d2) models on the light curves in Fig.~\ref{fig:Eff_min_max}.}
 \label{fig:eff_source}
\end{figure*}

\begin{figure*}
\centering
\includegraphics[width=\textwidth]{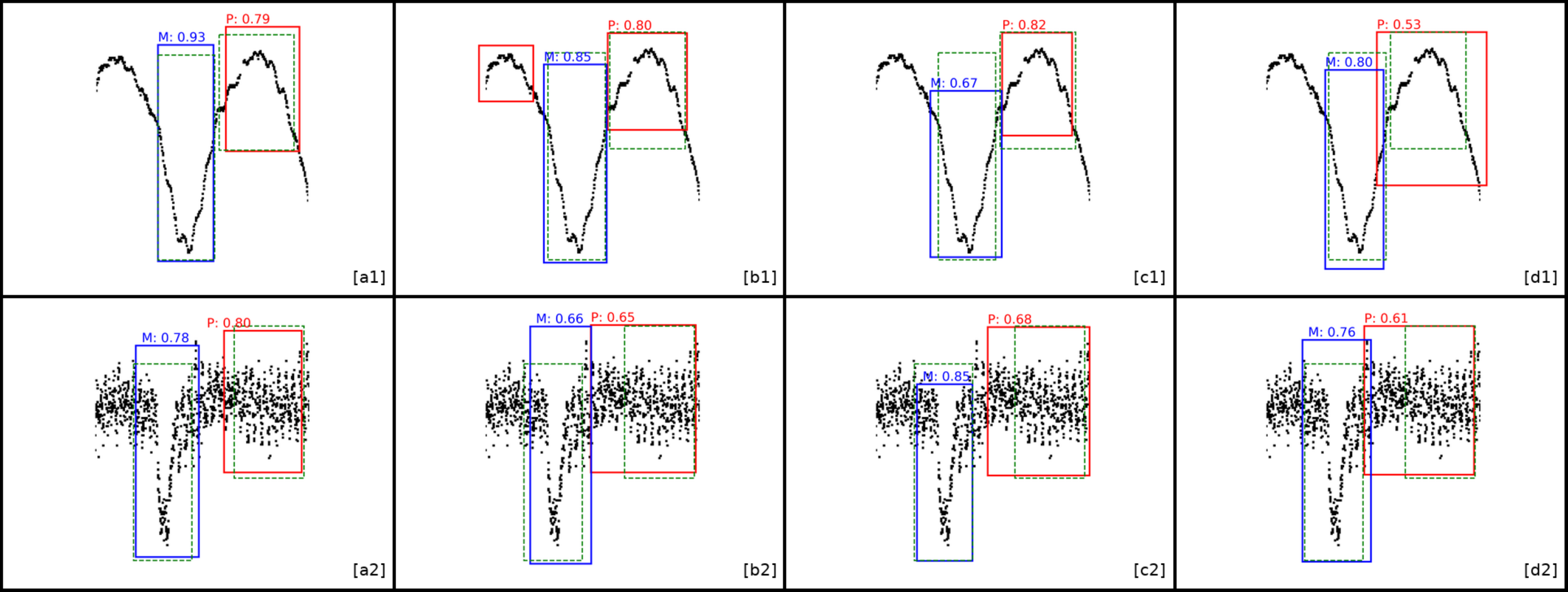} \\
\caption{Same as Fig.~\ref{fig:SSD_source}, but for the detection performance of SSD (a1, a2), Faster-RCNN (b1, b2),  YOLO (c1,c2)  and EfficientDet~D1 (d1, d2) models on the light curves in Fig.~\ref{fig:cnn_min_max}.}
 \label{fig:cnn_source}
\end{figure*}

We applied detections to the data with the highest and lowest IoUs from each model using the other four models (Figs. \ref{fig:SSD_source}-\ref{fig:cnn_source}) to provide a clear comparison of the results. In reviewing the 9 light curves used in this procedure, we found the overall average IoU values to be 0.83, 0.79, 0.76, 0.67, and 0.22 for the SSD, Faster R-CNN, YOLO, EfficientDet~D1, and non-pretrained CNN-based models, respectively. Although IoUs vary depending on the pattern occurrence within images, the pre-trained models clearly demonstrate their precision in this comparison.

\section{Test on {\it Kepler} data}\label{sec:keptest}

\begin{figure*}
\centering
\includegraphics[width=\textwidth]{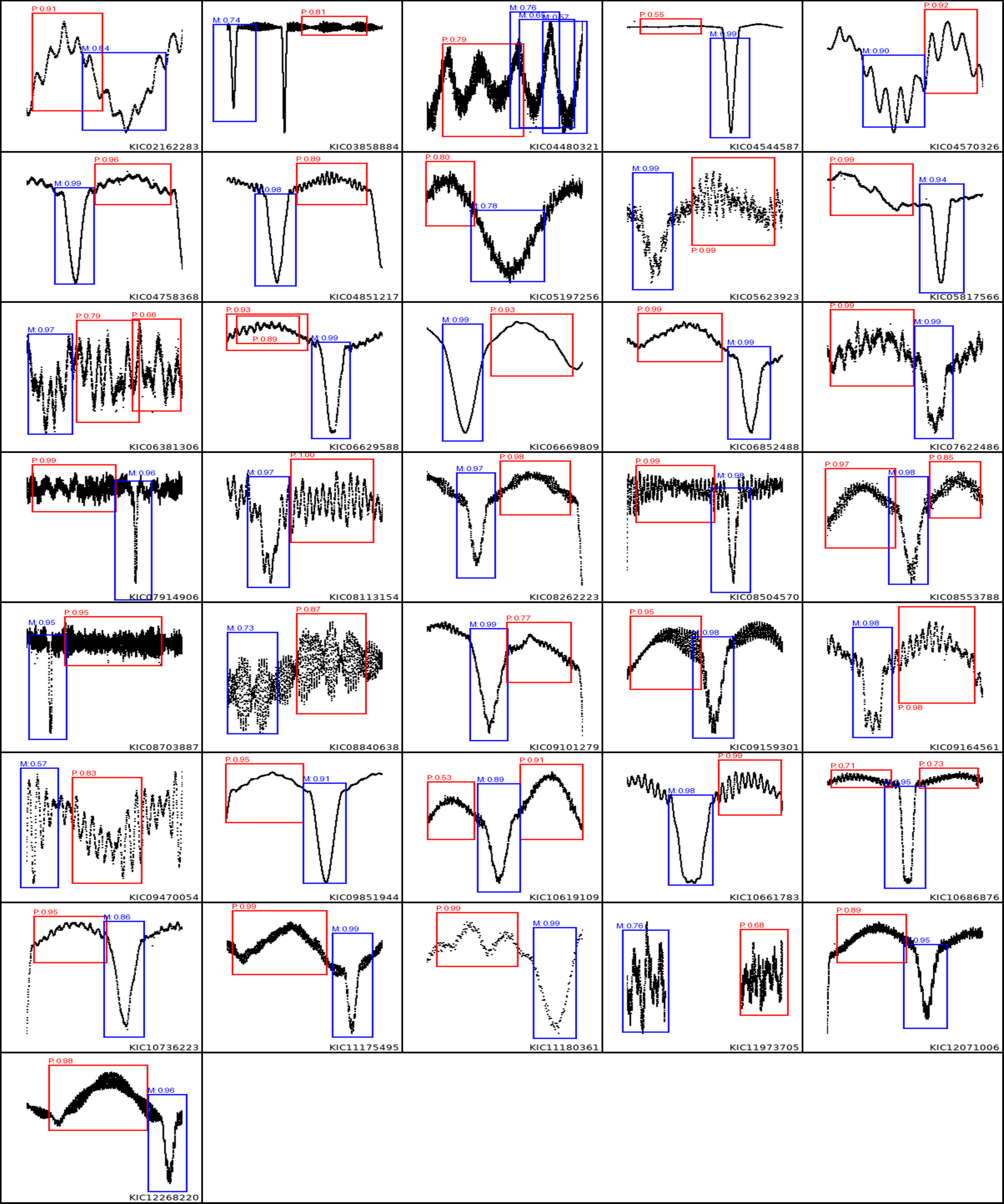} \\
\caption{Mosaic of 36 detections with the highest average confidence scores on short cadence {\it{Kepler}} data of 40 targets using the SSD model. The values above the red and blue bounding boxes refer to the confidence of pulsation (P) and minimum (M) patterns.}
 \label{fig:SSD_mosaic}
\end{figure*}

\begin{figure*}
\centering
\includegraphics[width=\textwidth]{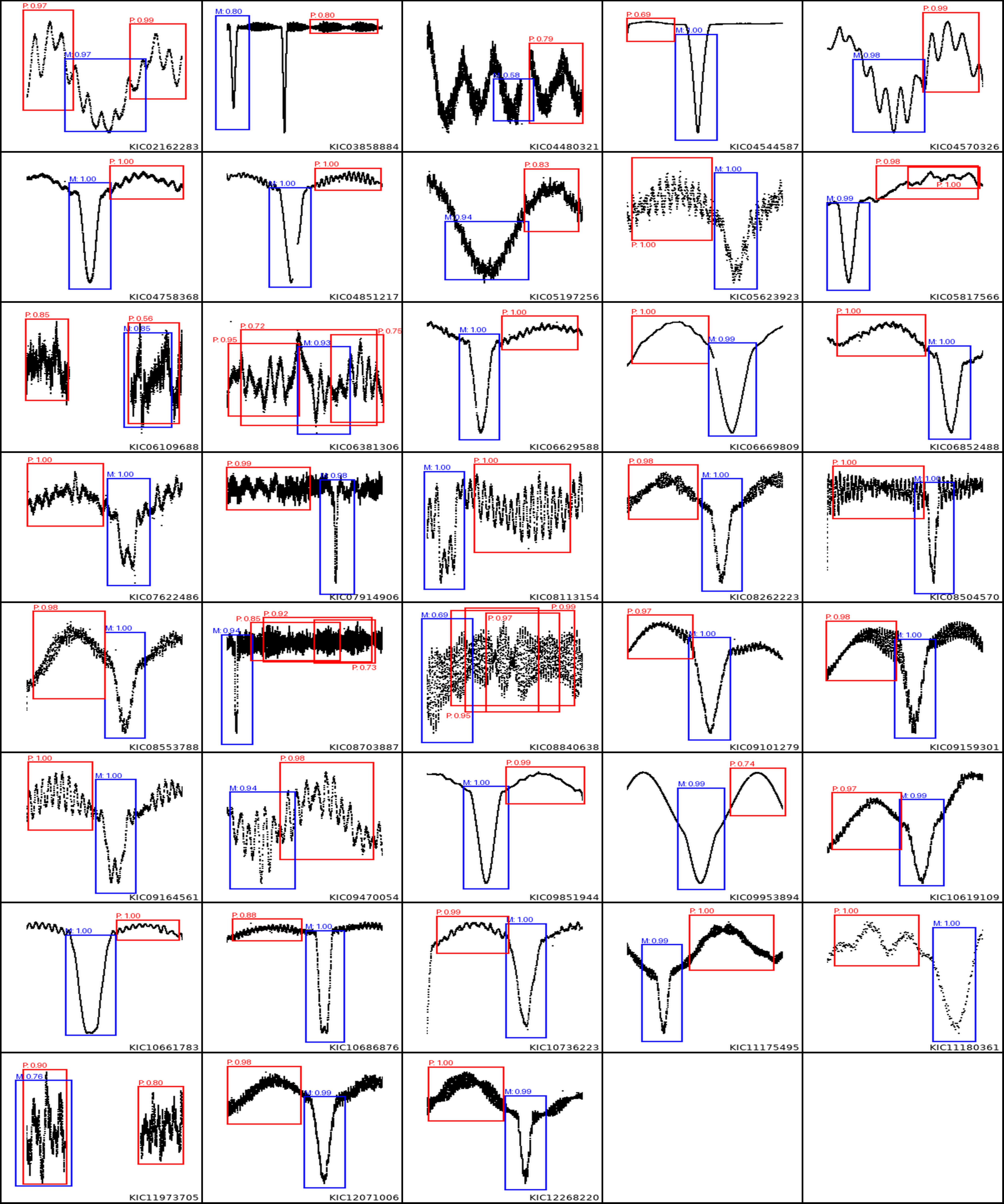} \\
\caption{Same as Fig.~\ref{fig:SSD_mosaic}, but for 38 detection using Faster R-CNN model.}
 \label{fig:frcnn_mosaic}
\end{figure*}

\begin{figure*}
\centering
\includegraphics[width=\textwidth]{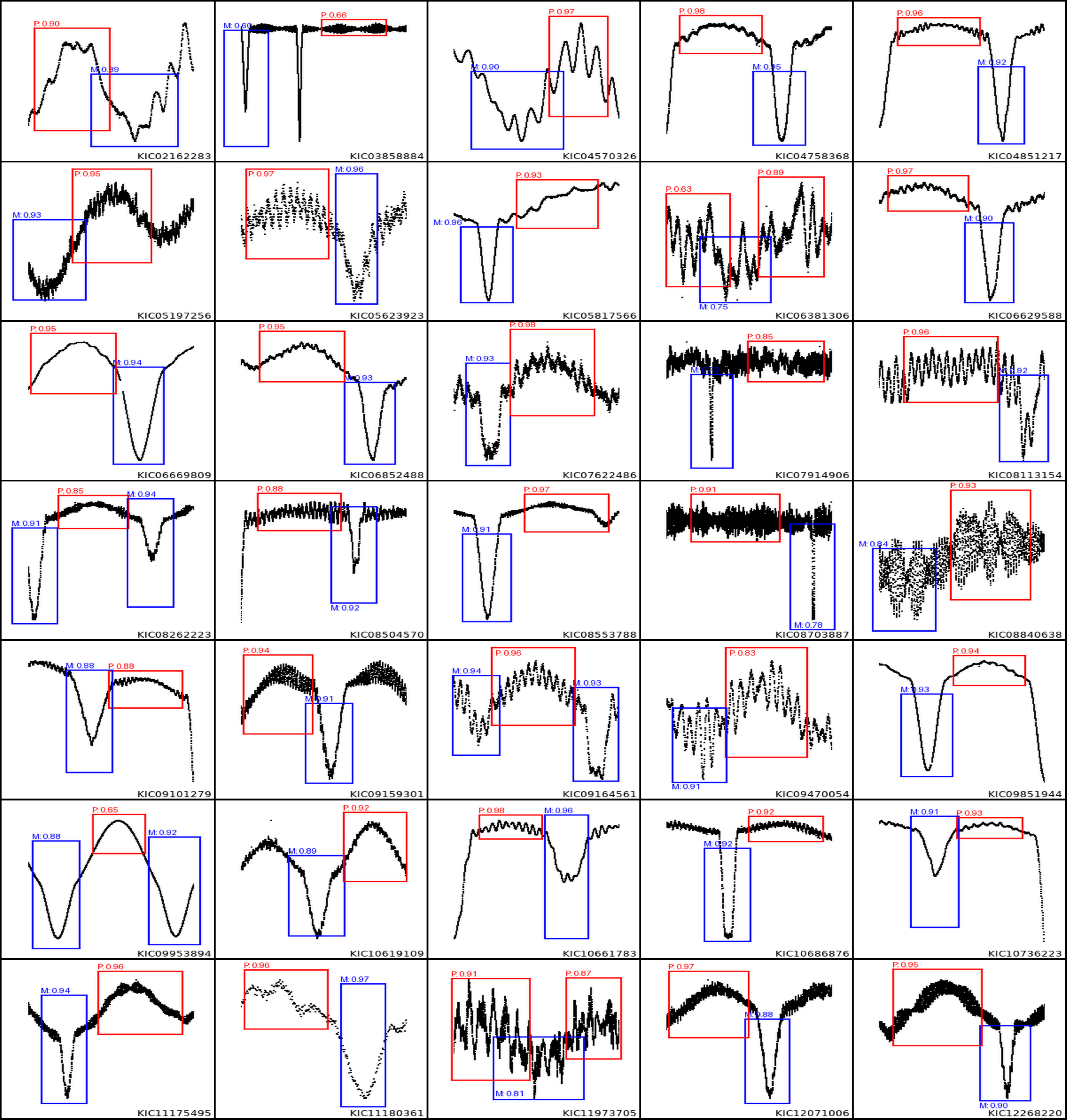} \\
\caption{Same as Fig.~\ref{fig:SSD_mosaic}, but for 35 detection using YOLO model.}
 \label{fig:yolo_mosaic}
\end{figure*}

The complete procedure mentioned above was conducted using TESS light curve data. It is, however, crucial to use  another database to test the model performance to the unseen data of different sources and check the usability with other datasets. In this manner, {\it{Kepler}} light curve data, especially taken in short cadence, are very appropriate to apply the test using the models with the best results, SSD, Faster R-CNN, and YOLO, out of the five of ours. We applied the test to the {\it{Kepler}} short cadence data of 124 systems with $\delta$ Sct type pulsating components, whose KIC numbers are given by \citet{lia17}. It must be emphasized that we did not cross-match the {\it{Kepler}} stars with our TESS targets, since the purpose is to test the inference performance using a different database and the number of known systems is rather low. Among the 124 systems, 40 have short-cadence data with the object ID in the FITS header matching the KIC numbers of the systems in the test set. The procedure described in Sec~\ref{sec:obsdata} was used to construct the light curve images. A modified version of \textsc{DetOcS}, \texttt{\it{{detocs\_k.py}}}\footref{fn:repo}, was used during the test with two models, SSD and Faster R-CNN. Testing on the {\it{Kepler}} dataset using the YOLO model, on the other hand, was conducted by employing the corresponding inference code\footref{fn:yol} . The confidence threshold was set to 0.5 during the procedures. Table~\ref{tab:keptest} lists the detection results showing the maximum average confidence values (sum of confidences divided by the number of detected patterns) for the systems with the model used in the detections where Faster R-CNN dominates with the highest values. Both pulsation and minimum patterns were detected in 36 targets out of 40 by SSD, 38 by Faster R-CNN and 35 by YOLO. Fig.~\ref{fig:SSD_mosaic}, \ref{fig:frcnn_mosaic} and \ref{fig:yolo_mosaic} illustrate the detections having the largest average confidence values for each target. The confidence scores displayed above the corresponding boxes in the in the light curves are designed to reflect the model’s certainty in its predictions. Bounding boxes with lower confidence scores, as seen in some detections such as KIC~4480321, KIC~4544587, and KIC~9470054 in SSD and Faster R-CNN models, indicate that the model is less certain of these predictions. These low-confidence results should be interpreted with caution, as they hold less significance compared to high-confidence predictions. The systems shown in the figures are those with the highest average scores, rather than the highest individual scores, as the selection process prioritized the average due to the presence of two classes of patterns. Another reason for focusing on detections with the highest average confidence scores is that we did not consider the total number of patterns or their locations. Detecting a single pulsation pattern along with the minimum, having a high confidence score - regardless of its position in the image - is sufficient to indicate that the system is a potential EBPC candidate. The different light curve shape of the same target for different models in the figures rises from the result that the models detect patterns with the highest confidence in the image constructed using different time intervals for a given target (see Sec.~\ref{sec:data}). 

\begin{table*}
    \caption{Maximum confidence scores for detected patterns on the light curves of 40 known EBPCs having short cadence {\it{Kepler}} data. Avg. refer average of confidence scores. P1, P2, P3, M1, M2 and M3 represent the number of patterns detected for pulsation and minimum.  The last column indicates the used model achieving the maximum average confidence, where FRCNN corresponds to the Faster R-CNN model.}
    \label{tab:keptest}
    \centering
    \begin{tabular}{lcccccccc}
    \hline
Target & P1 & P2 & P3 & M1 & M2 & M3 & Avg. & Model \\
\hline
KIC02162283 & 0.986 & 0.973 &  & 0.966 &  &  & 0.975 & FRCNN \\
KIC03858884 & 0.805 &  &  & 0.797 &  &  & 0.801 & FRCNN \\
KIC04480321 & 0.792 &  &  & 0.759 & 0.650 & 0.568 & 0.692 & SSD \\
KIC04544587 & 0.687 &  &  & 0.997 &  &  & 0.842 & FRCNN \\
KIC04570326 & 0.988 &  &  & 0.981 &  &  & 0.984 & FRCNN \\
KIC04758368 & 0.997 &  &  & 0.999 &  &  & 0.998 & FRCNN \\
KIC04851217 & 0.998 &  &  & 0.997 &  &  & 0.998 & FRCNN \\
KIC05197256 & 0.951 &  &  & 0.934 &  &  & 0.943 & YOLO \\
KIC05623923 & 0.999 &  &  & 0.999 &  &  & 0.999 & FRCNN \\
KIC05817566 & 0.997 & 0.976 &  & 0.989 &  &  & 0.988 & FRCNN \\
KIC06109688 & 0.855 & 0.565 &  & 0.849 &  &  & 0.756 & FRCNN \\
KIC06381306 & 0.950 & 0.752 & 0.721 & 0.932 &  &  & 0.839 & FRCNN \\
KIC06629588 & 0.998 &  &  & 0.998 &  &  & 0.998 & FRCNN \\
KIC06669809 & 0.996 &  &  & 0.994 &  &  & 0.995 & FRCNN \\
KIC06852488 & 0.996 &  &  & 0.996 &  &  & 0.996 & FRCNN \\
KIC07622486 & 0.999 &  &  & 0.998 &  &  & 0.999 & FRCNN \\
KIC07914906 & 0.986 &  &  & 0.975 &  &  & 0.980 & FRCNN \\
KIC08113154 & 0.997 &  &  & 0.998 &  &  & 0.997 & FRCNN \\
KIC08262223 & 0.982 &  &  & 0.998 &  &  & 0.990 & FRCNN \\
KIC08264510 & \multicolumn{8}{c}{\it{-No detections-}}   \\
KIC08504570 & 0.997 &  &  & 0.998 &  &  & 0.998 & FRCNN \\
KIC08553788 & 0.984 &  &  & 0.998 &  &  & 0.991 & FRCNN \\
KIC08703887 & 0.948 &  &  & 0.949 &  &  &  & SSD \\
KIC08840638 & 0.988 & 0.970 & 0.946 & 0.694 &  &  & 0.899 & FRCNN \\
KIC09101279 & 0.973 &  &  & 0.998 &  &  & 0.986 & FRCNN \\
KIC09159301 & 0.982 &  &  & 0.997 &  &  & 0.990 & FRCNN \\
KIC09164561 & 0.999 &  &  & 0.997 &  &  & 0.998 & FRCNN \\
KIC09470054 & 0.982 &  &  & 0.940 &  &  & 0.961 & FRCNN \\
KIC09851944 & 0.988 &  &  & 0.998 &  &  & 0.993 & FRCNN \\
KIC09953894 & 0.744 &  &  & 0.989 &  &  & 0.867 & FRCNN \\
KIC10619109 & 0.973 &  &  & 0.993 &  &  & 0.983 & FRCNN \\
KIC10661783 & 0.996 &  &  & 0.996 &  &  & 0.996 & FRCNN \\
KIC10686876 & 0.884 &  &  & 0.999 &  &  & 0.942 & FRCNN \\
KIC10736223 & 0.988 &  &  & 0.999 &  &  & 0.993 & FRCNN \\
KIC11175495 & 0.997 &  &  & 0.993 &  &  & 0.995 & FRCNN \\
KIC11180361 & 0.996 &  &  & 0.999 &  &  & 0.998 & FRCNN \\
KIC11973705 & 0.875 & 0.912 &  & 0.811 &  &  & 0.866 & YOLO \\
KIC12071006 & 0.979 &  &  & 0.995 &  &  & 0.987 & FRCNN \\
KIC12268220 & 0.996 &  &  & 0.987 &  &  & 0.991 & FRCNN \\
\hline
    \end{tabular}
\end{table*}

\begin{figure*}
\centering
\includegraphics[width=\textwidth]{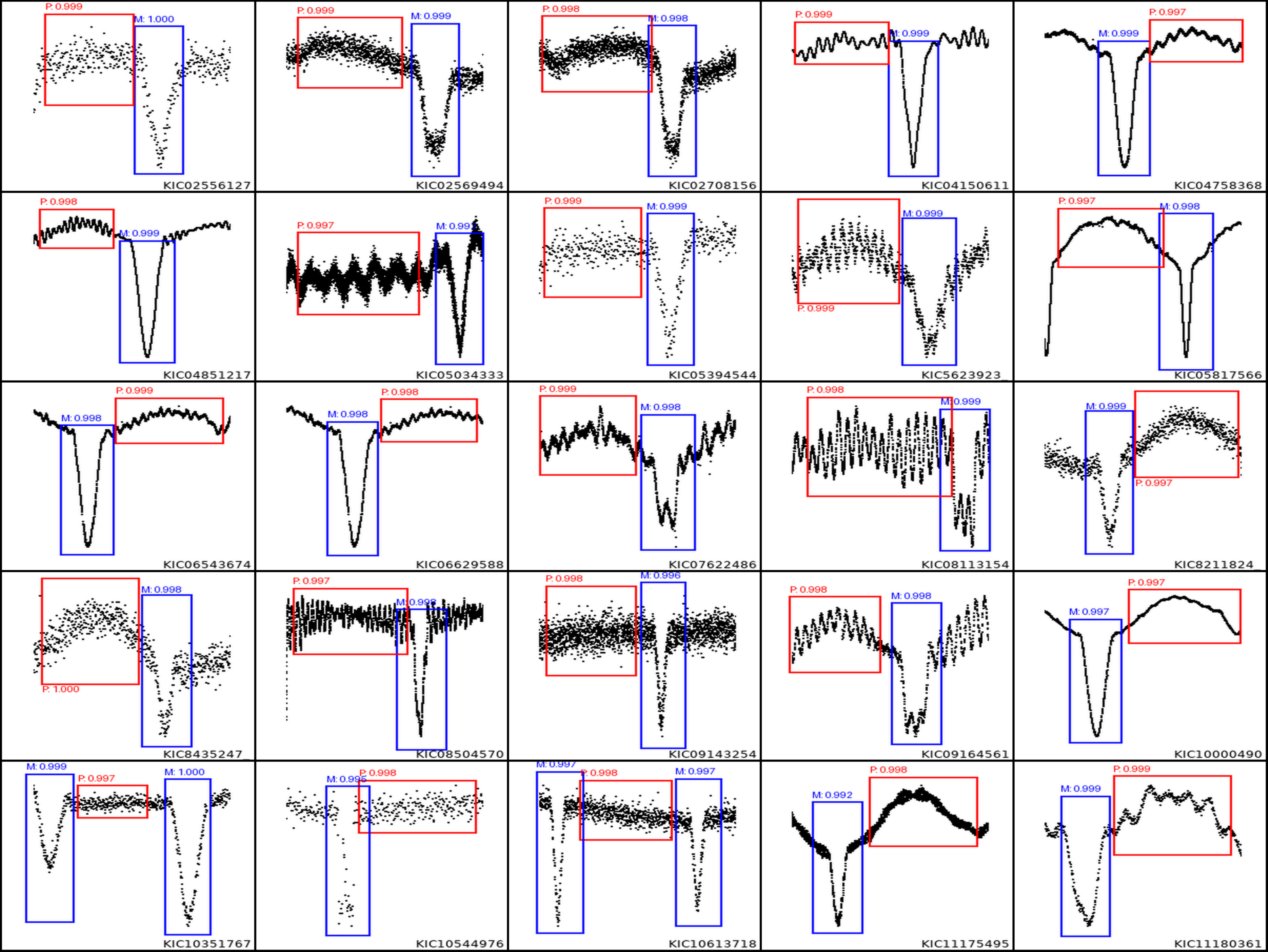} \\
\caption{Same as Fig.~\ref{fig:SSD_mosaic}, but for the light curves of the systems having the highest 25 confidence scores of pulsation patterns. Be aware that the inference results are contaminated by large scatterings in maximum phases of some light curves. See text for details.}
 \label{fig:kep25_mosaic}
\end{figure*}

We also attempted to apply the detection of pulsation and minimum patterns in the light curves of {\it{Kepler}} Eclipsing Binary systems listed in \citet{mat12}. Although the catalogue consists of several parameters besides orbital periods of 2920 systems, we managed to access short-cadence {\it{Kepler}} data of 574 binaries. The same detection procedure explained above was conducted, but on this occasion, only the Faster R-CNN was used, as it is the model that showed the highest confidence scores in our previous detection experience. Since many of the systems are not EBPCs in the catalogue, the model detected only minimum pattern(s) in the vast majority of the images. Fig.~\ref{fig:kep25_mosaic} displays the light curves that exhibit both pulsation and minimum patterns in eclipsing binaries with appropriate data from the catalog. For this particular figure, instead of using the average confidence score for pulsation and minimum patterns as is generally done, we have chosen to highlight the light curves with the 25 highest confidence scores for pulsation patterns in order to emphasize the most prominent cases of pulsation in the dataset. As seen in the figure, the model also treated certain scatterings as pulsations in some images. This behavior may arise due to the noise-like appearance of real pulsations in the light curves of some known EBPCs in our training set where the relatively high-frequency oscillations are hardly distinguishable in the selected time interval, that is determined based on the orbital period and includes the minimum pattern. Therefore, being on the safe side, those detections should not be considered pulsations without confirmation by further analysis. As an outline, the results indicate that the model is promising in eliminating the large amounts of light curves that are not showing two patterns together, namely not potentially EBPCs. It is also capable of detecting the patterns of interest correctly very fast, however, a final human inspection is still needed to confirm that a given system is an EBPC.

\section{Results and Conclusion}\label{sec:results}

We present the efforts to detect certain shapes similar to the effect of pulsation in eclipsing binary light curves by using conventional CNN-based object detection algorithms besides a custom one. The synthetic and observational light curve images forming our dataset were used to train the models and validate the performances. Table~\ref{tab:modsum} summarizes the key properties of the models used in this study, including the changes made to align with our objectives. Several codes, accessible via the internet\footref{fn:repo}, were written to conduct the procedures more autonomously. A web implementation, \textsc{DetOcS}, was also developed to apply inference based on certain models and to increase the number of samples in the dataset using the refinement method. It also allows the user to detect patterns of interest by entering a set of parameters, which makes it sustainable in discovering new candidates for future usage. 

\begin{table*}
\centering
\begin{threeparttable}
 \caption{Overview of object detection models used in the study. Layers refer to the total number of layers and parameters are the total number of trainable parameters. Speed corresponds to detection speed in frame per second. The values differ from the defaults as the models were modified to fit our task.}
    \label{tab:modsum}
\begin{tabular}{lccccc}
\hline
Model& Backbone   & Layers & Parameters &Speed (FPS) &Pretrained Dataset\\
\hline
SSD             & MobileNet v2                   & 324                   & 4627367 &    4.55                         & COCO 2017\\
Faster R-CNN    & ResNet50 V1                     & 275                   & 28278415 &   14.29                     & COCO 2017\\                  
YOLOv5          & CSP-Darknet                    & 270                   & 8923647   &   100                       & COCO 2017\\                   
EfficientDet D1 & EfficientNet B1                & 849                   & 4991810    &  1.72                       & COCO 2017\\                   
Non-pretrained CNN & custom Conv2D layers\tnote{a}   & 21                    & 810523  &    11.11                & -- \\                   
\hline
\end{tabular}
  \begin{tablenotes}
  \footnotesize
  \item[a] See Sec.~\ref{sec:npcnn}.
  \end{tablenotes}
\end{threeparttable}
\end{table*}

\begin{figure}
\centering
\includegraphics[width=0.5\textwidth]{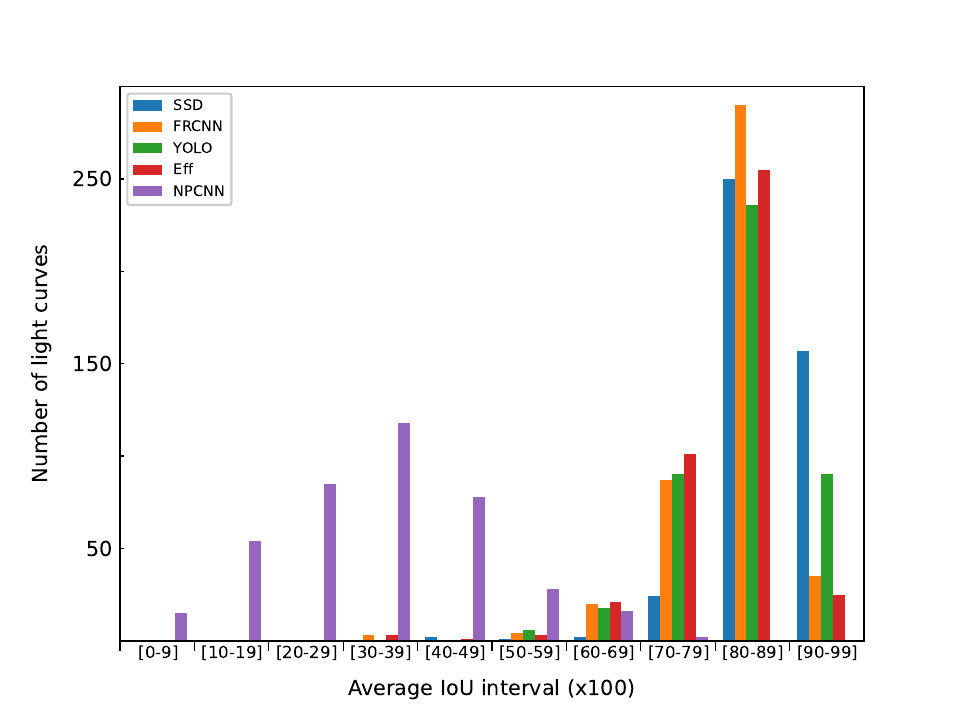} \\
\caption{A histogram showing the distribution of average IoU values achieved in detecting patterns on the observational light curves in which both classes (P and M) are detected. The numbers in the box brackets refer to the value intervals. Eff stands for EfficientDet~D1 model.}
 \label{fig:hist}
\end{figure}

\begin{figure*}
\centering
\includegraphics[width=\textwidth]{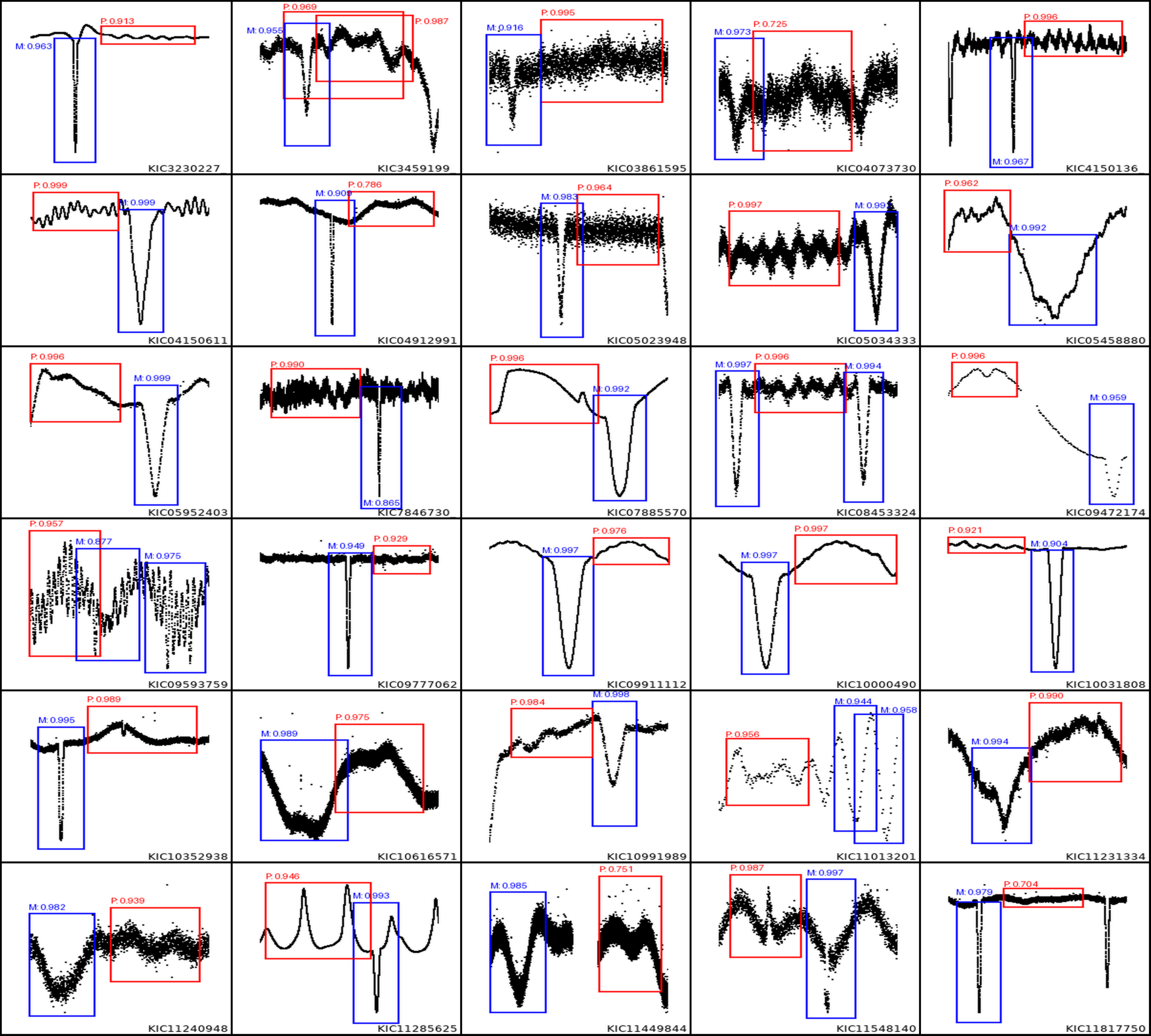} \\
\caption{Same as Fig.~\ref{fig:frcnn_mosaic}, but for the 30 systems in {\it{Kepler}} Eclipsing Binaries catalogue.}
 \label{fig:kepall_mosaic}
\end{figure*}

Our results imply that the Faster R-CNN shows the best performance to unseen data (Table~\ref{tab:keptest}) among the models introduced in the present study. The metrics obtained from SSD were also sufficient, which makes it useful considering its speed in both training and detecting patterns. YOLO is the fastest in detecting targets and resulted in a very good performance (Table~\ref{tab:metrics}), however, it was slow in the training process compared to others. The EfficientDet~D1 was the model showing lower performance in both training and inference in the manner of speed and accuracy, although the results are adequate as it correctly detected almost 90\% of the patterns on observational light curves in the validation set. In general, three of our models can be considered successful in detecting oscillation-like and minimum-shaped patterns in more than 94\% of the observational eclipsing binary light curve data. The custom CNN model is the weakest one which lets us consider that a non-pretrained model may not be powerful enough to achieve the desired numbers of successful detections. A histogram of the average IoU values obtained by the models is also plotted in Fig.~\ref{fig:hist}. The IoUs are considerably high and even exceed 0.9 for pre-trained models, SSD, Faster R-CNN, YOLO and EfficientDet~D1, while they are gathered in the region of the low values for the non-pretrained CNN-based model. The figure indicates that the values also peak at 0.8-0.9 interval for the four models. When we evaluate results concerning Precision-Recall curves, YOLO comes to the forefront in all IoU thresholds considering the consistency of the curve, as discussed in Sec.~\ref{sec:ssd}, where we explained the ideal properties of the curve, including maintaining high precision and recall across different decision thresholds. It is followed by the Faster R-CNN and SSD models with substantial trends, especially at IoU values of 0.5 and 0.75. Given the outcomes from the different phases of our study, the results clearly show that pre-trained conventional models work considerably better in the present task, detecting the patterns of interest with satisfactory results. A further detection with the Faster~R-CNN model was also applied to the light curves of the systems in {\it{Kepler}} Eclipsing Binaries having the short-cadence data. The results indicate that 68 systems show an oscillation-like pattern in their light curves after human-eye inspection of the inferences was conducted. Fig.~\ref{fig:kepall_mosaic}, a continuation of Fig.~\ref{fig:frcnn_mosaic}, shows the detected patterns in the light curves of 30 systems with the highest average confidence scores, excluding the 38 known EBPCs presented earlier. The variability types that may be responsible for the detected light variations, as provided in the literature, are listed in Table~\ref{tab:ext_kep}. To summarize, the performances of the models indicate that the object detection approach has the potential to scan the light curves and detect new eclipsing binary candidates with pulsating components in the present and future space- and ground-based missions. The data from different databases can be accessed and shaped in a suitable form almost autonomously using the series of codes we presented. The \textsc{DetOcS} can also be used to conduct detections in various databases with few modifications based on the accessed data source. The method is considerably fast in detecting patterns as the elapsed time for detection per image takes less than half of a second when using our model files, as mentioned in previous sections.

\begin{table*}
\centering
\begin{threeparttable}
 \caption{Possible causes of light variations observed in the {\it{Kepler}} targets in Fig.~\ref{fig:kepall_mosaic} based on previous studies given in Ref. column. $RG$ stands for Red Giant solar-like oscillator. Ell., Erp., Exo., Pul., Rot., Spo., Tid., Tri. correspond to ellipsoidal, eruption, exoplanet, pulsation, rotation, spot modulation, tidal induction, and triple system phenomenons, respectively.}
 \label{tab:ext_kep}
\begin{tabular}{lcclcc}
\hline
Target	&	Variability	&	Ref. & Target	&	Variability	&	Ref.\\
\hline
KIC03230227 & Tid. & 1  & KIC09593759 & Exo. & 15 \\
KIC03459199 & Pul. & 2  & KIC09777062 & Pul.($\gamma~Dor$). & 16 \\
KIC03861595 & Exo. & 3  & KIC09911112 & Pul.($\gamma~Dor$) & 17 \\
KIC04073730 & Erp. & 4  & KIC10000490 & Ell. & 7 \\
KIC04150136 & Tid. & 5  & KIC10031808 & Pul.($\gamma~Dor$) & 16 \\
KIC04150611 & Pul.($\delta~Sct$) & 6  & KIC10352938 & Exo. & 15 \\
KIC04912991 & Spo. & 7  & KIC10616571 & Rot. & 18 \\
KIC05023948 & Tri. & 8  & KIC10991989 & Pul.($RG$) & 16 \\
KIC05034333 & Tid. & 9  & KIC11013201 & Pul.($\delta~Sct$) & 19 \\
KIC05458880 & Rot. & 10  & KIC11231334 & Erp. & 14 \\
KIC05952403 & Tri. & 11  & KIC11240948 & Tid. & 4 \\
KIC07846730 & Spo. & 7  & KIC11285625 & Pul.($\gamma~Dor$) & 20 \\
KIC07885570 & Pul. & 12  & KIC11449844 & Exo. & 21 \\
KIC08453324 & Pul. & 13  & KIC11548140 & Erp. & 22 \\
KIC09472174 & Pul.($\beta~Cep$) & 14  & KIC11817750 & Pul.($\gamma~Dor$) & 17 \\
  
\hline
\end{tabular}
  \begin{tablenotes}
  \footnotesize
 \item  1.~\citet{guo17}, 2.~\citet{har04}, 3.~\citet{pat24}, 4.~\citet{bro21}, 5.~\citet{li23}, 6.~\citet{koe14}, 7.~\citet{lur17}, 8.~\citet{bre16}, 9.~\citet{li24}, 10.~\citet{bal15}, 11.~\citet{bor13}, 12.~\citet{shi22}, 13.~\citet{sch20}, 14.~\citet{deb11}, 15.~\citet{arm21}, 16.~\citet{gau19}, 17.~\citet{sek20}, 18.~\citet{ang18}, 19.~\citet{uyt11}, 20.~\citet{deb13}, 21.~\citet{mor16}, 22.~\citet{dav16}
  \end{tablenotes}
\end{threeparttable}
\end{table*}

As the following step of the present work, we plan to apply the trained object detection models on all available TESS and {\it{Kepler}} datasets to identify more EBPCs. Given these databases' huge quantity and complexity, this will be a challenging task that requires careful organization. However, our methods have demonstrated the ability to rapidly detect targets, even in relatively large and complex data. This efficiency will be particularly beneficial as we scale up to the full datasets, enabling us to infer potential targets much faster than traditional methods. We anticipate that optimizing model performance on this larger scale will involve iterative tuning of parameters, such as time intervals and confidence scores. Nonetheless, the code and methodologies have prepared us to handle these challenges with confidence. This extended application will validate our models, increase the number of potential EBPCs candidates, and uncover new insights within the datasets. Furthermore, more accurate data from future missions like PLATO \citep{rau14} will also provide a vast playground for applying/fine-tuning our models. Alternatively, similar object detection models can be trained for detecting other patterns or shapes corresponding to a variety of astrophysical phenomenons to discover new targets of interest for in-depth examination. 

\begin{acknowledgements}
BU acknowledges the financial support of TUBITAK (The Scientific and Technological Research Council of Turkey) within the framework of the 2219 International Postdoctoral Research Fellowship Program (No.1059B192202496). Hospitalitiy of the HUN-REN CSFK Konkoly Observatory is greatly appreciated. RSz and TSz acknowledge the support of the SNN-147362 grant of the  Hungarian  Research,  Development and  Innovation  Office  (NKFIH). The numerical calculations reported in this paper were partially performed at TUBITAK ULAKBIM, High Performance and Grid Computing Center (TRUBA resources). This paper includes data collected with the TESS mission, obtained from the MAST data archive at the Space Telescope Science Institute (STScI). Funding for the TESS mission is provided by the NASA Explorer Program. STScI is operated by the Association of Universities for Research in Astronomy, Inc., under NASA contract NAS 5–26555. This paper includes data collected by the Kepler mission and obtained from the MAST data archive at the Space Telescope Science Institute (STScI). Funding for the Kepler mission is provided by the NASA Science Mission Directorate. STScI is operated by the Association of Universities for Research in Astronomy, Inc., under NASA contract NAS 5–26555. This work made use of Astropy:\footnote{\url{http://www.astropy.org}} a community-developed core Python package and an ecosystem of tools and resources for astronomy \citep{ast13, ast18, ast22}. This research has made use of the SIMBAD database, operated at CDS, Strasbourg, France. The authors acknowledge OpenAI's ChatGPT for its assistance in debugging certain codes used in this research, with all suggested corrections carefully reviewed and validated by the authors before implementation.

\end{acknowledgements}

\bibliographystyle{aa} 
\bibliography{ref} 

\begin{appendix}
\section{}\label{sec:appa}
We present the formulae and assumptions used in parameter determination for eclipsing binaries during the construction of synthetic light curves. The catalogs mention in Sec.~\ref{subsec:par_det} \citep{sou15,mal20} lists the effective temperatures ($T_{e_1}$, $T_{e_2}$), orbital period of the systems ($P$), masses ($M_{1}$, $M_{2}$), and radii ($R_{1}$, $R_{2}$) of the components. 

The mass ratio ($q$) was calculated using the formula $q=\frac{M_{2}}{M_{1}}$ by assuming that the $M_{1}$ is more massive than $M_{2}$ for semidetached binaries. 

The surface potential of the primary component was determined using the modified Kopal potential \citep{kop59}, under the assumption of synchronous rotation:
\[
\Omega_1 = \frac{1}{r_1} + q \left(\frac{1}{(r_1^2 - 2r_1 + 1)^{\frac{1}{2}}} - r_1\lambda\right) + \frac{1}{2}(q + 1)(1-\nu^2) r_1^2
\]
where $r_1$ is the fractional radius, $r_1 = \frac{R_1}{a}$, with $a$ being the separation calculated using Kepler’s third law, $M_1 + M_2 = \frac{a^3}{P^2}$, and the corresponding assumptions. Furthermore, $\lambda = \sin{\theta} \cos{\phi}$ and $\nu = \cos{\theta}$, where $\theta$ and $\phi$ are the polar and azimuthal angles, respectively. $\Omega_1$ was derived for doublets of ($\lambda, \nu$) and remains constant over the surface.

The surface potential of the secondary component was then derived by:
\[
\Omega_2 = \frac{\Omega_1}{q} + \frac{q-1}{2q}
\]

The inclination, $i$, was determined under the assumption that for an eclipse to occur the projected separation ($a\sin{i}$) must be less than or equal to the sum of the radii of the stars, $a\sin{i} \leqslant  R_1 + R_2$ and therefore, $\sin{i} \leqslant \frac{R_1 + R_2}{a}$ which corresponds to the following using $\sin{^2}{i} + \cos^{2}{i} = 1$:
\[
\cos{i} \geq \sqrt{1-\frac{R_1 + R_2}{a}}.
\]
Considering the inclination is between $\left [ 0^{\circ}, 90^{\circ} \right ]$ and thus the positive root, the following can be written:
\[
i \geq \cos^{-1}({\frac{R_1 + R_2}{a}})
\]
which was set as the minimum value in our study. It was increased following the method explained in Sec.~\ref{subsec:par_det}.

\section{}\label{sec:appb}

We provide additional information about the specifications, the operational steps of models and the key parameters/hyperparameters necessary for their successful execution and satisfactory performance in the following subsections.

\subsection{Single Shot Multibox Detector Model}

The steps involved in the SSD detection process as follows. The SSD model begins by accepting an input image, which it resizes to a certain size and normalizes for consistency. It uses a backbone network, MobileNet v2 in our case, to extract feature maps from the image. These feature maps represent the input at multiple scales, enabling the detection of objects of varying sizes. SSD divides the feature map into a grid and assigns each cell multiple default bounding boxes with varying aspect ratios. Each default box predicts a confidence score for every class and refines its location. Finally, SSD applies Non-Maximum Suppression \citep[NMS,][]{red16} to remove overlapping boxes with lower confidence scores, resulting in the final bounding boxes, confidence scores, and class labels.

The model contains a predefined anchor box used to estimate the bounding boxes and class probabilities. The most important feature of the SSD is the capability to set a balance between detailed spatial patterns and high-level semantic patterns, which lets the algorithm detect objects with diverse characteristics. The relatively fast estimations of the method spread its usage for various aims like object detection in complex scenarios. Our model was SSD algorithm with MobileNet backbone which is pre-trained on the Common Objects in Context 2017 \citep[COCO,][]{lin14}. It combines the efficiency of MobileNet~v2 architecture with the accuracy of the SSD framework. Its structure is compact and requires low computational power. Working on moderate resolution sets the model to balance speed and precision. The data record files, label maps and a pipeline configuration file, which are crucial for training, were created using the API’s related codes. 

Several series of \texttt{batch\_size}, \texttt{warmup\_learning\_rate} and \texttt{learning\_rate\_base} parameters were examined by considering the commonly used values for SSD models. Eventually, our final model in re-training yielded by fine-tuning these values as 32, 0.0001 and 0.0005, respectively. The L$_2$ regularization \citep{rum86} was also added by setting its weight to 0.00006. The different step size values were also tried. We concluded that the model in the 6200th epoch is the most optimal value for the parameter (\texttt{num\_step}) to avoid overfitting in the retraining process as there was no significant improvement in validation loss and its difference from the training loss during the next steps of training.  The complete process was conducted on the Malor server of Konkoly Observatory using one Nvidia GeForce RTX 2080 Ti accelerator. The loss function was selected as a combination of classification and localization losses which are sigmoid and smooth L1 loss functions, along with the regularization loss. USing the final model it took 771.75$\,\mathrm{seconds}$ to detect patterns on 3411 validation images.

\subsection{Faster R-CNN Model}

Faster R-CNN starts by processing an input image of arbitrary size, normalizing its pixel values before feeding it into a backbone network \citep[ResNet50 V1,][]{lin14}. The backbone extracts feature maps, which are passed to the Region Proposal Network (RPN). The RPN scans the feature maps to generate region proposals, identifying areas likely to contain objects. These proposals are then aligned with the feature map using ROI (Region of Interest) pooling, producing fixed-size feature maps for each region. Finally, these feature maps are used to classify the object in each region and refine the bounding box coordinates. The model outputs the final bounding boxes, confidence scores, and class labels. 


The model was pre-trained on the COCO 2017 dataset and accessible through the internet\footref{fn:tfzoo}. The TensorFlow Object Detection API was used during the process, similar to done in the previous section. \texttt{batch\_size} was set to 4 which dispersed the loss curve slightly, however, no memory overflow was observed during training in return. In the final model, \texttt{warmup\_learning\_rate} was equal to 0.0001 until the 2000th epochs while \texttt{learning\_rate\_base} was adapted to 0.001. One Nvidia GeForce RTX 2080 Ti accelerator was used during the procedure.

\subsection{YOLO Model}

The YOLO detection process operates as follows. The model divides the input image into a grid, resizes it, and normalizes its pixel values. Each grid cell is responsible for predicting bounding boxes and their associated class probabilities. YOLO employs a deep convolutional neural network to extract features and capture relationships between objects. It performs object localization, classification, and confidence score prediction in a single forward pass. Each prediction includes bounding box coordinates, a confidence score indicating the likelihood of an object, and probabilities for all possible classes. Finally, NMS is applied to remove overlapping boxes, resulting in the final bounding boxes, confidence scores, and class labels.

Since the model needs fine-tuning to detect our specific objects of interest, similar to previous ones, the size was modified by setting the parameters of the \texttt{depth\_multiple} and \texttt{width\_multiple} to 0.5. The \texttt{learning\_rate} was 0.001 while the weight decay (\texttt{weight\_decay}) of the Stochastic Gradient Descent \citep{lec15} optimizer was adopted to 0.001, a larger value from the default. The only non-zero augmentation parameter was image translation (\texttt{translate} = 0.1), besides default HSV augmentations. A PyTorch-implemented version of the YOLO was used by aiming a leverage in GPU acceleration and automatic differentiation. The training procedure was applied using 2 x Nvidia GeForce RTX 2080 Ti accelerator.

\subsection{EfficientDet D1 Model}

EfficientDet-D1 processes an input image by resizing it to match the model's requirements and normalizing its pixel values. It uses EfficientNet (specifically the B1 variant in our case) as its backbone for feature extraction, followed by the Bidirectional Feature Pyramid Network \citep[BiFPN,][]{tan20}, which merges multi-scale features to improve detection performance. Anchor boxes are generated at multiple scales and aspect ratios, enabling the model to predict objects of different sizes. Each anchor box predicts class probabilities and bounding box offsets. EfficientDet-D1 applies NMS to filter overlapping predictions, producing the final bounding boxes, confidence scores, and class labels. The D1 sub-model has increased detection performance in terms of balance between model size and speed.

The training was made using TensorFlow Object Detection API and related files in the zoo\footref{fn:tfzoo}. The first train process using a pre-trained model showed no improvement in loss functions beyond the 10000th epoch (33$\,\mathrm{minutes}$ 32$\,\mathrm{seconds}$) and resulted in a small number of correctly detected samples in the validation dataset based on the model from the last epoch. Therefore, we retrained the model using the last obtained checkpoint from the initial training. To avoid the extreme load on GPU memory and fail in memory allocation, we set the batch\_size parameter to 4. A learning rate scheduler was set by adopting \texttt{warmup\_learning\_rate} and \texttt{learning\_rate\_base} parameters to 0.0001 and 0.001, respectively, while the weight of the L$_2$ regularizer was adopted to 0.00004. The same accelerator, one Nvidia GeForce RTX 2080 Ti, was used during training and evaluation processes. 

\subsection{A non-pretrained CNN-based Model}
The model mainly consists of a convolutional part for extracting features, a dense head to convert feature maps into a 2D vector, three fully connected layers for classification, box regression, and determining confidence. The feature extraction has three convolutional layers with the 16, 16 and 32 filters beside L$_2$ regularizer in the first one. The network consists of 17 hidden layers, with 12 originating from the feature extraction component and 5 from the dense processing stage. The {\tt{Adam}} \citep{kin15} was used as an optimizer with an initial learning rate of 0.0005, although a learning rate scheduler ({\tt{ReduceLROnPlateau}}) was applied to reduce it when validation loss stopped improving. Early stopping callback was also employed based on the validation loss to prevent overfitting, hence, it stopped the training process at the 97th epoch since the value did not improve in the last ten epochs. 

\end{appendix}

%
%
\end{document}